\documentclass{amsart}
\usepackage{fullpage,amssymb,amsmath,amsthm,amsfonts}
\usepackage{ae,aecompl}
\usepackage{ifthen}
\usepackage{graphicx}
\usepackage{url}
\newboolean{dum}
\setboolean{dum}{false}
\newcommand{\comment}[1]{\ifthenelse{\boolean{dum}}{
{\par\noindent\Huge\ding{46}} \fbox{\parbox{10cm}{#1}}\par}{}}

\newtheorem{lemma}[subsection]{Lemma}
\newtheorem{claim}[subsection]{Claim}
\newtheorem{thm}[subsection]{Theorem}
\newtheorem{prop}[subsection]{Proposition}
\newtheorem{cor}[subsection]{Corollary}

\newtheorem{defs}[subsection]{Definition}
\numberwithin{equation}{section}

\newcommand{\lr}{\left [}
\newcommand{\rr}{\right ]}

\begin{document}
\title{Tridiagonal realization of the anti-symmetric Gaussian $\beta$-ensemble}
\author{Ioana Dumitriu}
\address{
Department of Mathematics, 
University of Washington, 
Seattle, USA}
\author{Peter J. Forrester} 
\address{Department of Mathematics and Statistics, 
The University of Melbourne,  
Victoria 3010, Australia}
\begin{abstract}{The Householder reduction of a member of the anti-symmetric Gaussian
unitary ensemble gives an anti-symmetric tridiagonal matrix with all independent
elements. The random variables permit
the introduction of a positive parameter $\beta$, and the eigenvalue
probability density function of the corresponding random matrices can be computed
explicitly, as can the distribution of $\{q_i\}$, the first components of 
the eigenvectors.
Three proofs are given. One involves an inductive construction based on 
bordering of a family of random matrices which are shown to have the same
distributions as the anti-symmetric tridiagonal matrices. This proof uses the
Dixon-Anderson integral from Selberg integral theory. A second proof involves the 
explicit computation of the Jacobian for the change of variables between real anti-symmetric tridiagonal matrices,
its eigenvalues and $\{q_i\}$. The third proof maps
matrices from the anti-symmetric Gaussian $\beta$-ensemble to those realizing
particular examples of the Laguerre $\beta$-ensemble. In addition to these proofs, we note some simple properties of
the shooting eigenvector and associated Pr\"ufer phases of the random matrices.}
\end{abstract}
\maketitle

\section{Introduction}

Gaussian ensembles of random matrices are best known for their application 
to quantum mechanics. The aim of this application (see e.g.~\cite{Ha90}) 
is to predict the statistical properties of highly excited energy levels 
when the underlying classical mechanics is chaotic; toward this purpose, 
the Hamiltonian is modelled by a random Hermitian matrix $H$. Time 
reversal symmetry requires that the elements of $H$ be real, while there 
being no preferential basis in determining the spectrum requires that the 
joint probability density function (PDF) of $H$ be invariant under 
conjugation by orthogonal transformations.

These constraints are all satisfied by the Gaussian orthogonal ensemble (GOE), which consists of real random matrices $H = {1 \over 2} (X + X^T)$, where $X$ is an $n \times n$ Gaussian matrix with all entries independent standard normals. The joint distribution of all independent entries is thus seen to be proportional to $\exp ( - {\rm Tr} \, H^2/2)$, which is invariant under conjugation by orthogonal transformations, $H \mapsto O H O^T$. This key property of the GOE explains the adjective ``orthogonal'' in its name.

In the theory of quantum conductance through a  normal metal -- superconductor junction
the matrix modelling the Hamiltonian must have a block structure to account for
both electrons and holes. In the case when there is no time reversal 
symmetry,
nor spin-rotation invariance, this block matrix must be of the form \cite{AZ97}
$$
\begin{bmatrix} A & B \\
- \bar{B} & - \bar{A} \end{bmatrix}, \qquad A = A^\dagger, \quad B = - B^T.
$$
Such matrices are equivalent under conjugation to $i$ times a real
anti-symmetric matrix, and so motivate the consideration of anti-symmetric
Gaussian matrices $\tilde{H} =
{i \over 2} (\tilde{X} - \tilde{X}^T)$ where $\tilde{X}$ is an
$N \times N$ real Gaussian matrix with entries ${\rm N}[0,1/2]$.
Here and throughout the paper ${\rm N}[\mu,\sigma^2]$ refers to the normal 
distribution with mean $\mu$
and variance $\sigma^2$. This class of random matrices --- referred to as
the anti-symmetric Gaussian unitary ensemble --- has also received recent attention for
its appearance in the study of point processes relating 
to the tiling of the half hexagon by three species of rhombi, and to
classical complex Lie algebras \cite{FN08a,De08a}.

It is our objective in this paper to initiate a study of anti-symmetric Gaussian matrices
from the viewpoint of the underlying tridiagonal matrices. To appreciate the possible
scope for such a study, let us recall
that
the GOE has a ``sibling'' ensemble of tridiagonal matrices which shares 
the same eigenvalue PDF. The latter was constructed by applying a well-known numerical algorithm for eigenvalue computation to the GOE. For a general $n \times n$ real symmetric matrix, a common preliminary strategy in numerical eigenvalue computation is to first conjugate by a sequence of Householder reflection matrices so as to obtain a similar tridiagonal matrix. In the case of GOE matrices, it was shown by Trotter \cite{Tr84} that the resulting tridiagonal matrix exhibits a remarkable property: all elements are again independently distributed (subject only to symmetry). Explicitly, one obtains tridiagonal matrices
\begin{equation}\label{25}
\left [ \begin{array}{ccccc} {\rm N}[0,1] & \tilde{\chi}_{n-1} & & & \\
\tilde{\chi}_{n-1} & {\rm N}[0,1] & \tilde{\chi}_{n-2} & & \\
 & \tilde{\chi}_{n-2} & {\rm N}[0,1] & \tilde{\chi}_{n-3} & \\
& \ddots & \ddots & \ddots & \\
&&\tilde{\chi}_{2} & {\rm N}[0,1] & \tilde{\chi}_{1} \\
& & & \tilde{\chi}_{1} & {\rm N}[0,1] \end{array} \right ]
\end{equation}
where
${\rm N}[0,1]$ refer to the standard normal distribution
and $\tilde{\chi}_{k}$
denotes the square root of the  gamma distribution
$\Gamma[k/2,1]$, the latter  being
specified by
the p.d.f.~$(1/ \Gamma(k/2))
u^{k/2-1} e^{-u}$, $u > 0$. 

Two other Gaussian ensembles also apply to the highly excited states of chaotic
quantum systems, namely
the Gaussian unitary ensemble (GUE) of complex Hermitian matrices and the
Gaussian symplectic ensemble (GSE) of Hermitian matrices with real
quaternion entries (see e.g.~\cite{Fo02}). It was pointed out by
Dumitriu and Edelman \cite{DE02} that applying the Householder transformation
to these matrices gives the tridiagonal matrix (\ref{25}) but with the
replacements 
\begin{equation}\label{3.1}
\tilde{\chi}_{k} \mapsto \tilde{\chi}_{\beta k} \qquad (k=1,\dots,n-1)
\end{equation}
where $\beta = 2$ for the GUE and $\beta = 4$ for the GSE. Moreover, with the
replacements (\ref{3.1}) for general $\beta > 0$ it was proved in
\cite{DE02} that the eigenvalue PDF of the corresponding random tridiagonal
matrices is equal to
\begin{equation}\label{3.2}
{1 \over \tilde{G}_{\beta, n}}
\prod_{i=1}^n e^{-\lambda_i^2/2} \prod_{1 \le j < k \le n}
| \lambda_k - \lambda_j |^\beta
\end{equation}
where
$$ 
\tilde{G}_{\beta, n} = (2 \pi)^{n/2} \prod_{j=0}^{n-1}
{\Gamma(1 + (j+1)\beta/2) \over \Gamma(1 + \beta/2)}.
$$

Our specific aim then is to give random tridiagonal matrices
whose  eigenvalue
PDF generalizes that of the anti-symmetric Gaussian matrices (as well
as the ensemble consisting of pure imaginary elements, another ensemble 
that can be constructed out of pure imaginary real quaternion elements).
As in (\ref{25}), we
find that the tridiagonal matrices in question have independent elements (up to the anti-symmetry condition), and that these elements allow for a
$\beta$ generalization. Explicitly,
our study focuses on 
the family of random anti-symmetric tridiagonal matrices
\begin{equation} \label{3.3}
A_{n}^{\beta} = \left [ \begin{array}{ccccc} 0 & \tilde{\chi}_{(n-1)\beta/2} & & & \\
-\tilde{\chi}_{(n-1)\beta/2} & 0 & \tilde{\chi}_{(n-2)\beta/2} & & \\
 & - \tilde{\chi}_{(n-2)\beta/2} & 0 & \tilde{\chi}_{(n-3)\beta/2} & \\
& \ddots & \ddots & \ddots & \\
&&-\tilde{\chi}_{\beta} & 0 & \tilde{\chi}_{\beta/2} \\
& & & - \tilde{\chi}_{\beta/2} & 0 \end{array} \right ] ~.
\end{equation}
For $n$ even, the eigenvalues of these matrices come in pairs $\{ \pm i \lambda_j \}_{j=1,\dots,n/2}$,
$\lambda_j > 0$, while for $n$ odd zero is a simple eigenvalue, and the remaining
eigenvalues come in pairs $\{ \pm i \lambda_j \}_{j=1,\dots,(n-1)/2}$,
$\lambda_j > 0$. 

Our main results are given in Theorem \ref{one} 
and Theorem \ref{two}, 
for which we give three proofs; each proof uses different tools and highlights different properties of the matrices in question. 

\begin{thm} \label{one}
With the notation above, the PDF of the positive variables
$\{\lambda_j\}$, ordered non-decreasingly ($\lambda_1 \leq \lambda_2 \leq \ldots \leq \lambda_n$), for $n$ even is given by
\begin{equation}\label{pn1}
\frac{1}{C_{\beta,n}} ~ \prod_{i=1}^{n/2} \lambda_i^{\beta/2 - 1} e^{- \lambda_i^2}
\prod_{1 \le j < k \le n/2} (\lambda_j^2 - \lambda_k^2)^{\beta}~,
\end{equation}
while for $n$ odd it is given by 
\begin{equation} \label{pn2}
\frac{1}{C_{\beta,n}} ~ \prod_{i=1}^{(n-1)/2} \lambda_i^{3\beta/2 - 1}  e^{- \lambda_i^ 2}
\prod_{1 \le j < k \le (n-1)/2} (\lambda_j^2 - \lambda_k^2)^{\beta}~.
\end{equation}
The normalization constants are given by
\begin{eqnarray} \label{z1}
\left \{ \begin{array}{ll} \displaystyle
C_{\beta,n} = \prod_{j=1}^{n/2} {\Gamma(2j\beta/4) \Gamma((2j-1)\beta/4) \over 2 \Gamma(\beta/2) }, & n \: \: {\rm even} \\ \displaystyle
\\
C_{\beta,n} = {\Gamma(n\beta/4) \over \Gamma(\beta/4) } C_{(n-1),\beta}, &
n \: \: {\rm odd} \end{array} \right.
\end{eqnarray}
\end{thm}

These PDFs are related to $\beta$-Laguerre ensembles (see \cite{Fo97}, \cite{DE02}). More precisely, using the notation of \cite{DE02}, the squares of the eigenvalues of antisymmetric Gaussian $\beta$-ensembles of size $n$ are distributed like the eigenvalues of $\beta$-Laguerre ensembles of size $\lfloor \frac{n}{2} \rfloor$ and parameters $a = \frac{(n-1)\beta}{4}$ for $n$ even, and $a = \frac{n\beta}{4}$ for $n$ odd. 

In addition to the PDFs, we give the distributions of the first components of the eigenvectors. 

\begin{thm} \label{two}
Let $q_j$, $j=1,\dots, \lfloor \frac{n}{2} \rfloor$ be the (positive) 
first component of the independent eigenvector corresponding to $\{\lambda_j\}$. In addition, if $n$ is odd, let $z$ be the (positive) first component of the eigenvector corresponding to the (simple) $0$ eigenvalue. Then 
\begin{itemize}
\item if $n$ is even, the vector $(q_1, \ldots, q_{n/2})$ has Dirichlet distribution $D_{n/2}[(\beta/2)^{n/2}]$ (here $(\beta/2)^{n/2}$ denotes $\beta/2$ repeated $n/2$ times);
\item if $n$ is odd, the vector $(q_1, \ldots, q_{(n-1)/2}, z)$ has Dirichlet distribution $D_{(n-1)/2}[(\beta/2)^{(n-1)/2},\beta/4]$.
\end{itemize}
\end{thm}
 
The rest of the paper is organized as follows. In Section \ref{house_red} we begin by recalling the definition of the two classical ensembles of anti-symmetric Gaussian matrices, and show that the Householder reduction of the anti-symmetric Gaussian unitary ensemble gives (\ref{3.3}) with $\beta = 2$. Although we do not give details, the  Householder reduction in the case of
pure imaginary real quaternion elements gives (\ref{3.3}) with $\beta = 4$. 

We show three different approaches to the eigenvalue PDF computation for the matrix model \eqref{3.3} with general $\beta$; all of them extend to  the computation of the PDF of the first component of the corresponding eigenvectors. Each proof is self-contained. The first one, in Sections \ref{peter_proof} and \ref{peter_proof2},   
uses the characteristic polynomial of the matrix, which is constructed inductively. A second approach, shown in 
Section \ref{ioana_proof1}, is based on Jacobians of tridiagonal anti-symmetric matrix. A third proof is presented in Section \ref{ioana_proof2}, where we compute the PDF of Theorem \ref{one} by use of an 
orthogonal similarity transformation between the anti-symmetric $\beta$-ensemble and a $\beta$-Laguerre 
ensemble.
Finally, in Section \ref{apps} we make note of some simple properties of
the shooting eigenvector and associated Pr\"ufer phases
of the random matrices (\ref{3.3}).

\section{Householder reduction} \label{house_red}
As mentioned, there are two classical ensembles of anti-symmetric
Gaussian matrices. One is the anti-symmetric Gaussian unitary ensemble, which 
is specified in the paragraph below (\ref{3.2}). The other is the anti-symmetric
of interest in statistical physics might have $\beta$-generalizations; one case in point is that of the antisymmetric Gaussian of interest in statistical physics might have $\beta$-generalizations; one case in point is that of the antisymmetric Gaussian of interest in statistical physics might have $\beta$-generalizations; one case in point is that of the antisymmetric Gaussian of interest in statistical physics might have $\beta$-generalizations; one case in point is that of the antisymmetric Gaussian of interest in statistical physics might have $\beta$-generalizations; one case in point is that of the antisymmetric Gaussian Gaussian symplectic ensemble. A member $A$ of this ensemble is formed out of
an $n \times n$ Gaussian random matrix $Y$ with quaternion real entries
$$
Y_{ij} = \left [ \begin{array}{cc} z & w \\ - \bar{w} & \bar{z} \end{array}
\right ],
$$
(as a complex matrix, $Y$ is $2n \times 2n$), according to
$A = {1 \over 2}(Y - Y^T)$. Both the $z$ and $w$ variables
are required to be purely imaginary, chosen from $i$N[0,1]. This ensemble
has the special feature that the eigenvalues are doubly degenerate.
Generally the eigenvalues of anti-symmetric Hermitian matrices come in pairs
$\{\pm i \lambda_j \}_{j=1,\dots,[n/2]}$ $(\lambda_j > 0)$ and with zero as a simple eigenvalue for $n$ odd. 

The PDFs of the positive eigenvalues for both the anti-symmetric
GUE and GSE are known (see e.g.~\cite{Fo02}). For the anti-symmetric GUE the PDF is proportional to
\begin{equation}\label{gp1}
\left \{ \begin{array}{ll} \displaystyle
\prod_{i=1}^{n/2}  e^{- \lambda_i^2}
\prod_{1 \le j < k \le n/2} (\lambda_j^2 - \lambda_k^2)^{2}, &
n \: \: {\rm even} \\ \displaystyle
\prod_{i=1}^{(n-1)/2} \lambda_i^{2}  e^{- \lambda_i^2}
\prod_{1 \le j < k \le (n-1)/2} (\lambda_j^2 - \lambda_k^2)^{2}, &
n \: \: {\rm odd} \end{array} \right.
\end{equation}
while for the anti-symmetric GSE it is proportional to
\begin{equation}\label{gp2}
\left \{ \begin{array}{ll} \displaystyle
\prod_{i=1}^{n/2}  \lambda_i e^{- \lambda_i^2}
\prod_{1 \le j < k \le n/2} (\lambda_j^2 - \lambda_k^2)^{4}, &
n \: \: {\rm even} \\ \displaystyle
\prod_{i=1}^{(n-1)/2} \lambda_i^{5}  e^{- \lambda_i^2}
\prod_{1 \le j < k \le (n-1)/2} (\lambda_j^2 - \lambda_k^2)^{4}, &
n \: \: {\rm odd}. \end{array} \right.
\end{equation}
Here we give the explicit form of the Householder reduction of anti-symmetric
GUE matrices, and so deduce a random tridiagonal matrix with eigenvalue PDF
proportional to (\ref{gp1}).

As for real symmetric matrices \cite{Wi65}, in the case of anti-symmetric matrices
$[a_{ij}]_{i,j=1,\dots,n}$, the Householder tridiagonalization consists of
constructing a sequence of symmetric real orthogonal matrices $O^{(j)}$
$(j=1,\dots,n-2)$ such that
$$
O^{(n-2)} O^{(n-3)} \cdots O^{(1)} A O^{(1)} O^{(2)} \cdots O^{(n-2)} = B^{(n-2)},
$$
where $ B^{(n-2)}$ is an anti-symmetric tridiagonal matrix. 

This process is based on the fact that any column vector $x= (x_1, \ldots, 
x_n)^T$ can be mapped into $||x||_2 e_1$, with $e_1$ being the first 
column of the identity matrix ${\mathbb I}_n$, by the symmetric orthogonal 
transformation 
$H = {\mathbb I}_n - 2 \frac{v \cdot v'}{||v||^2}$, where $v = x + x_1 
e_1$ and $||v|| :=||v||_2$. The 
matrix $H$ is called the \emph{Householder reflector} for $x$.

For our antisymmetric matrices, the matrix $O^{(1)}$ is chosen so that $B^{(1)} :=  O^{(1)} A  O^{(1)}$ is tridiagonal with respect to the first row and first column; the matrix $O^{(2)}$ is chosen so that $B^{(2)} :=
O^{(2)} B^{(1)} O^{(2)} $ is tridiagonal with respect to the first two rows and
first two columns, etc..~For this $O^{(j)}$ must be of the form
\begin{equation}\label{OR}
O^{(j)} = \left [ \begin{array}{cc} {\mathbb I}_j & {\mathbb O}_{j \times n-j} \\
 {\mathbb O}_{n-j \times j} & R^{(n-j)} \end{array} \right ]
\end{equation}
where $R^{(n-j)}$ is the Householder reflector corresponding to the $j+1$st through $n$th entries in the $j$th column of $O^{(j-1)} \cdots O^{(1)} A O^{(1)} \cdots O^{(j-1)}$. Here ${\mathbb O}_{i \times j}$ is the $i \times j$ matrix of all zero entries.

It follows that 
\begin{equation}\label{ba}
(B^{(1)})_{11} = a_{11}, \qquad (B^{(1)})_{12} = - (B^{(1)})_{21} =
(a_{12}^2 + \cdots a_{1n}^2)^{1/2}~,
\end{equation}
etc..

With real anti-symmetric matrices formed by $-i$ times an anti-symmetric GUE matrix, it
follows from (\ref{ba}) that $(B^{(1)})_{12}$ has distribution equal to the square root
of $\Gamma[(n-1)/2,1]$. 

Anti-symmetric GUE matrices are distributionally invariant under conjugation by an (independent) orthogonal matrix; this is an immediate consequence of the fact that a column vector of i.i.d. Gaussians does not change its distribution when left multiplied by an (independent) orthogonal matrix. 

This invariance under conjugation, as well as the structure (\ref{OR}) of $O^{(1)}$, tell us that the sub-block of $B^{(1)}$ obtained by deleting the first row and first column is $-i$ times an anti-symmetric GUE
matrix of size $n-1$. Proceeding inductively and remembering the structure (\ref{OR}),
we see that (\ref{3.3}) holds in the case $\beta = 2$.

\section{Method I part (i): bordered matrices and an inductive construction} \label{peter_proof}
An inductive construction of the tridiagonal matrices (\ref{25}), (\ref{3.1})
involving the operations of bordering was given in \cite{FR02b}.
In \cite{FN08a} this same bordering procedure was used to compute the joint
eigenvalue PDF of an anti-symmetric GUE matrix and its successive principal
minors. Here the working of \cite{FN08a} will be extended to provide an 
inductive construction of a family of random matrices with eigenvalue 
PDFs given by Theorem \ref{one}. Additional workings from \cite{FR02b} will then
be adapted to deduce the tridiagonal
matrix (\ref{25}) itself.

\begin{prop}\label{p3.1}
Let $\mathbf{w}$ be an $n$-component real column vector of the form
\begin{equation}\label{ww}
\mathbf{w} = \left \{
\begin{array}{ll} (w_1,w_1,\dots,w_{n/2},w_{n/2}), & n \: \: {\rm even} \\
(w_1,w_1,\dots,w_{(n-1)/2},w_{(n-1)/2},b), & n \: \: {\rm odd} \end{array} \right.
\end{equation}
Let $2 w_j^2$ have distribution $\Gamma[\beta/2,1]$ and let $b^2$ have distribution
$\Gamma[\beta/4,1]$. Define the sequence of Hermitian random matrices $A_1=0$,
$A_2$, $A_3, \dots$, where each $A_k$ is $k \times k$, by the inductive formula
\begin{equation}\label{Ak}
A_{n+1} = \left [
\begin{array}{cc} {\rm diag} \, A_n & i \mathbf{w} \\
- i \mathbf{w}^T & 0 \end{array} \right ]
\end{equation}
where ${\rm diag} \, A_n$ is the diagonal matrix formed from the eigenvalues of $A_n$.
The eigenvalues of each $A_k$ come in plus/minus pairs
 $\{\pm \lambda_j\}_{j=1,\dots,[k/2]}$, $\lambda_j > 0$, and for
$k$ odd zero is also a simple eigenvalue. (Such pairing is to be taken as implicit
in ${\rm diag} \, A_n$, and when $n$ is odd the zero eigenvalue is to be
listed last.) Furthermore,
with the characteristic polynomial of $A_n$ denoted by $P_n(x)$,
for $n$ even we have
\begin{equation}\label{w1}
{P_{n+1}(x) \over x P_n(x) } = 1 - \sum_{i=1}^{n/2} {2 w_i^2 \over x^2 - \lambda_i^2}~,
\end{equation}
while for $n$ odd
\begin{equation}\label{w2}
{P_{n+1}(x) \over x P_n(x) } = 1 - {b^2 \over x^2} -  \sum_{i=1}^{(n-1)/2} {2 w_i^2 \over x^2 - \lambda_i^2}.
\end{equation}
\end{prop}

\begin{proof} $\quad$ We see that if $\mathbf{v}$ is an eigenvector of (\ref{Ak}) with eigenvalue
$\lambda$, then $\bar{\mathbf{v}}$ is an eigenvector with
eigenvalue $-\lambda$.  
Thus, as claimed, the eigenvalues come in plus/minus pairs, and for $k$ odd
there is a zero eigenvalue corresponding to an eigenvector with all components real.
To establish the equations for the characteristic polynomial, we first note that by induction we must have 
\begin{equation}\label{3.5}
{\rm diag} \, A_n = \left \{
\begin{array}{ll} (\lambda_1 , - \lambda_1 ,\dots, \lambda_{n/2} , -\lambda_{n/2} ), & n \: \: {\rm even} \\
(\lambda_1,-\lambda_1,\dots,\lambda_{(n-1)/2},-\lambda_{(n-1)/2},0), & n \: \: {\rm odd} \end{array} \right.
\end{equation}
Recalling (\ref{ww}), it follows from (\ref{Ak}) that for $n$ even
$$
\det ( x \mathbb I_{n+1} - A_{n+1} ) =
\det ( x \mathbb I_{n} - A_{n} ) \Big ( x - \sum_{j=1}^{n/2} w_j^2 \Big (
{1 \over x - \lambda_j} + {1 \over x + \lambda_j} \Big ) \Big )
$$
while for $n$ odd
$$
\det ( x \mathbb I_{n+1} - A_{n+1} ) =
\det ( x \mathbb I_{n} - A_{n} ) \Big ( x - {b^2 \over x} - \sum_{j=1}^{(n-1)/2} w_j^2 \Big (
{1 \over x - \lambda_j} + {1 \over x + \lambda_j} \Big ) \Big ).
$$
These are the equations (\ref{w1}) and (\ref{w2}) respectively. \end{proof}

\medskip
The equations (\ref{w1}), (\ref{w2}) can be used to compute the conditional
eigenvalue PDF for $A_{n+1}$, given the eigenvalues of $A_n$. We see that the task
is to compute the density of the zeros of the random rational functions therein.
This can be done by appealing to the following known result.

\begin{prop} 
\cite[Corollary 3]{FR02b}
Consider the random rational function
\begin{equation}\label{R}
R(x) = 1 - \sum\limits_{i=1}^n {q_i \over x - a_i}
\end{equation}
where each $q_i$ has distribution $\Gamma[s_i,1]$. This function has exactly $n$ roots,
each of which is real, and for given $\{a_i\}$ these roots have PDF
$$
{1 \over \Gamma(s_1) \cdots \Gamma(s_n) }
e^{- \sum\limits_{j=1}^n (x_j - a_j) }
\prod_{1 \le i < j \le n} {(x_i - x_j) \over (a_i - a_j)^{s_i + s_j - 1} }
\prod_{i,j=1}^n | x_i - a_j|^{s_j - 1},
$$
supported on
$$
x_1 > a_1 > \cdots > x_n > a_n.
$$
\end{prop}

In the case of $n$ even, the RHS of (\ref{w1}) corresponds to (\ref{R}) with
$n \mapsto n/2$, $x \mapsto x^2$, $a_i \mapsto \lambda_i^2$ and each
$q_i$ distributed as $\Gamma[\beta/2,1]$. In the case $n$ odd, the RHS of
(\ref{w2}) corresponds to (\ref{R}) with $n \mapsto (n+1)/2$, $x \mapsto x^2$,
$a_i \mapsto \lambda_i^2$ $(i=1,\dots, (n-1)/2)$, $a_{(n+1)/2} = 0$, $q_i$
distributed as $\Gamma[\beta/2,1]$ $(i=1,\dots,(n-1)/2)$ and $q_{(n+1)/2}$
distributed as $\Gamma[\beta/4,1]$. The 
sought conditional PDFs can
therefore be made explicit.

\begin{prop}
For $n$ even, the PDF of the positive eigenvalues of $A_{n+1}$, given that the
positive eigenvalues of $A_n$ are $\lambda_1,\dots,\lambda_{n/2}$, is equal to
\begin{equation}\label{x1}
{2^{n/2} \prod\limits_{j=1}^{n/2} x_j e^{ - \sum\limits_{j=1}^{n/2} (x_j^2 - \lambda_j^2) } \over
( \Gamma(\beta/2))^{n/2} }
\prod_{1 \le i < j \le n/2}
{(x_i^2 - x_j^2) \over (\lambda_i^2 - \lambda_j^2)^{\beta - 1} }
\prod\limits_{i,j=1}^{n/2} | x_i^2 - \lambda_j^2 |^{\beta/2 - 1}
\end{equation}
where
\begin{equation}\label{Re1}
x_1 > \lambda_1 > \cdots > x_{n/2} > \lambda_{n/2}.
\end{equation}
For $n$ odd, the PDF of the positive eigenvalues of $A_{n+1}$, given that the
positive eigenvalues of $A_n$ are $\lambda_1,\dots,\lambda_{(n-1)/2}$, is
equal to
\begin{eqnarray}\label{x2}
&&{2^{(n+1)/2} e^{ - x_{(n+1)/2}^2 - \sum\limits_{j=1}^{(n-1)/2} (x_j^2 - \lambda_j^2)} \over
( \Gamma(\beta/2))^{(n-1)/2} \Gamma(\beta/4)} \nonumber \\
&& \qquad \times
{\prod\limits_{i=1}^{(n+1)/2} x_i^{\beta/2 - 1} \over
\prod\limits_{i=1}^{(n-1)/2} (\lambda_i^2)^{(3\beta/4 - 1)} }
~{\prod\limits_{{\tiny 1 \le i < j \le (n+1)/2}}
(x_i^2 - x_j^2) \over \prod\limits_{{\tiny 1 \le i < j \le (n-1)/2}}^{~} (\lambda_i^2 - \lambda_j^2)^{\beta - 1} }
\prod_{i=1}^{(n+1)/2} ~\prod_{j=1}^{(n-1)/2} | x_i^2 - \lambda_j^2 |^{\beta/2 - 1}
\end{eqnarray}
where
\begin{equation}\label{Re2}
x_1 > \lambda_1 > \cdots > x_{(n-1)/2} > \lambda_{(n-1)/2} > x_{(n+1)/2} > 0.
\end{equation}
\end{prop}

Let the conditional PDFs of the above proposition be denoted
$$
G_{n+1}(\{x_j\}_{j=1,\dots,[(n+1)/2]}; \{\lambda_j \}_{j=1,\dots,[n/2]}),
$$
and let the marginal PDF of the positive eigenvalues of $A_n$ be denoted
$p_n(\{x_j\}_{j=1,\dots,[n/2]})$. Then we must have that
\begin{eqnarray}\label{iq}
&& p_{n+1}(\{x_j\}_{j=1,\dots,[(n+1)/2]}) \nonumber \\
&& \quad =
\int_0^\infty d \lambda_1 \cdots \int_0^\infty d \lambda_{[n/2]} \,
G_{n+1}(\{x_j\}_{j=1,\dots,[(n+1)/2]}; \{\lambda_j \}_{j=1,\dots,[n/2]})
p_n(\{\lambda_j\}_{j=1,\dots,[n/2]}).
\end{eqnarray}
Furthermore, from the definition of $A_2$ we have that
\begin{equation}\label{ic}
p_2(x) = {2 \over \Gamma(\beta/4) } x^{\beta/2 - 1} e^{- x^2}, \qquad x>0
\end{equation}
so (\ref{iq}) uniquely specifies $\{p_n(\{x_j\}_{j=1,\dots,[n/2]}) \}_{n=3,4,\dots}$.
As such, it can be used to verify that the explicit functional form of $p_n$ is given by Theorem \ref{one}.


\vspace{.25cm}

\noindent
\underline{{\it {\bf First proof of Theorem \ref{one}.}}} We make a trial functional form
$$
p_n(\{\lambda_j\}_{j=1,\dots,[n/2]})
 = {1 \over C_{n,\beta} }
\prod_{i=1}^{[n/2]} e^{- \lambda_i^2} \lambda_i^{\alpha_{n,\beta}}
\prod_{1 \le j < k \le [n/2]} (\lambda_j^2 - \lambda_k^2)^\beta
$$
where $C_{n,\beta}$ and $\alpha = \alpha_{n,\beta}$ are to be determined.
Substituting in (\ref{iq}) then gives, for $n$ even,
\begin{eqnarray}\label{a3}
&& {C_{n,\beta} \over C_{n+1,\beta} } (\Gamma(\beta/2))^{n/2}
\prod_{i=1}^{n/2} x_i^{ \alpha_{n+1,\beta} - 1}
\prod_{1 \le j < k \le n/2} (x_j^2 - x_k^2)^{\beta - 1} \nonumber \\
&& \qquad =
2^{n/2} \int_{R_{n/2}} d \lambda_1 \cdots d \lambda_{n/2} \,
\prod_{k=1}^{n/2} \lambda_k^{\alpha_{n,\beta}}
\prod_{i < j}^{n/2} (\lambda_i^2 - \lambda_j^2)
\prod_{i,j=1}^{n/2} | x_i^2 - \lambda_j^2|^{\beta/2 - 1}
\end{eqnarray}
where $R_{n/2}$ is the region (\ref{Re1}), while for $n$ odd
\begin{eqnarray}\label{a4}
&& {C_{n,\beta} \over 2 C_{n+1,\beta} } ~(\Gamma(\beta/2))^{(n-1)/2} \Gamma(\beta/4)
\prod_{i=1}^{(n+1)/2} x_i^{ \alpha_{n+1,\beta} + 1 - \beta/2}
\nonumber \\ && \quad \times 
\prod_{i=1}^{(n+1)/2} x_i^{ \alpha_{n+1,\beta} + 1 - \beta/2}
\prod_{1 \le j < k \le (n+1)/2} (x_j^2 - x_k^2)^{\beta - 1} 
=
2^{(n-1)/2} \int_{R_{(n-1)/2}'} d \lambda_1 \cdots d \lambda_{(n-1)/2} \nonumber \\
&&\quad \times 
\prod_{l=k}^{(n-1)/2} \lambda_k^{\alpha_{n,\beta}+2-3\beta/2}
\prod_{i < j}^{(n-1)/2} (\lambda_i^2 - \lambda_j^2)
\prod_{i=1}^{(n+1)/2} \prod_{j=1}^{(n-1)/2} | x_i^2 - \lambda_j^2|^{\beta/2 - 1}
\end{eqnarray}
where $R_{(n-1)/2}'$ denotes the region of integration (\ref{Re2}).

To evaluate the integrals, we make use of the Dixon-Anderson integral
(see e.g.~\cite[Ch.~3]{Fo02})
\begin{eqnarray}\label{DA}
&&\int_X  d \lambda_1 \cdots d \lambda_n \, \prod_{1 \le j < k \le n}
(\lambda_j - \lambda_k) \prod_{j=1}^n \prod_{p=1}^{n+1} |\lambda_j - a_p|^{s_p - 1}
\nonumber \\
&& \qquad = {\prod\limits_{i=1}^{n+1} \Gamma(s_i) \over \Gamma \left (\sum\limits_{i=1}^{n+1} s_i \right ) }
\prod_{1 \le j < k \le n + 1} (a_j - a_k)^{s_j + s_k - 1}
\end{eqnarray}
where $X$ is the domain of integration
$$
a_1 > \lambda_1 > a_2 > \lambda_2 > \cdots > \lambda_n > a_{n+1}.
$$
After the simple change of variables 
$\lambda_j \mapsto \lambda_j^2$ and the replacements $a_j \mapsto x_j^2$, we see that for
\begin{equation}\label{a51}
 \alpha_{n,\beta} = 3\beta/2 - 1 \qquad (n \: \: {\rm odd})
\end{equation}
the RHS of (\ref{a4}) is equal to
$$
{\Gamma(\beta/2))^{(n+1)/2} \over \Gamma((n+1)\beta/4)}
\prod_{j<k}^{(n+1)/2} (x_j^2 - x_k^2)^{\beta - 1}.
$$
The fact that the above is identical to the LHS of \eqref{a4} provides
\begin{equation}\label{a5a}
{C_{n,\beta} \over 2 C_{n+1,\beta} }
{\Gamma(\beta/4) \Gamma((n+1)\beta/4) \over \Gamma(\beta/2)} = 1
\qquad (n \: \: {\rm odd})
\end{equation}
and
\begin{equation}\label{a52}
\alpha_{n+1,\beta} = \beta/2 - 1 \qquad (n \; \: {\rm odd})~.
\end{equation}
Furthermore, (\ref{DA}) with $a_{n+1} = 0$ allows us to compute the RHS of (\ref{a3}) as
equal to
$$
{\Gamma((\alpha_{n,\beta} + 1)/2) (\Gamma(\beta/2))^{n/2} \over
\Gamma(n \beta/4 + (\alpha_{n,\beta} + 1)/2) }
\prod_{i=1}^{n/2} x_i^{\beta - 1 + \alpha_{n,\beta} }
\prod_{j < k}^{n/2} (x_j^2 - x_k^2)^{\beta - 1}.
$$
This is identical to the LHS provided that 
\begin{equation}\label{a6a}
{C_{n,\beta} \over C_{n+1,\beta} }
{\Gamma(n \beta/4 + (\alpha_{n,\beta} + 1)/2) \over
\Gamma((\alpha_{n,\beta} + 1)/2) } = 1 \qquad (n \: \: {\rm even})
\end{equation}
and
\begin{equation}\label{a6}
\alpha_{n+1,\beta} = \alpha_{n,\beta} + \beta \qquad (n \: \: {\rm even}).
\end{equation}

We observe that (\ref{a51}), (\ref{a52}) and (\ref{a6}) overdetermine
$\alpha_{n,\beta}$, as (\ref{a51}) and  (\ref{a52}) determine $\alpha_{n,\beta}$
in all cases. Substituting (\ref{a51}) into the LHS and (\ref{a52}) into the
RHS of (\ref{a6}) shows this latter equation to be identically satisfied.
Further, we observe that (\ref{a52}) is consistent with the requirement of the
initial condition (\ref{ic}). With $\alpha_{n,\beta}$ thus specified,
substituting in (\ref{a6a}) simplifies that formula to read
\begin{equation}\label{a7}
{C_{n,\beta} \over C_{n+1,\beta}}
{\Gamma((n+1)\beta/4) \over \Gamma(\beta/4) } = 1 \qquad
(n \: \: {\rm even}).
\end{equation}
The equations (\ref{a5a}), (\ref{a7}) together with the requirement $C_{2,\beta} =
\Gamma(\beta/4)/2$ as read off from (\ref{ic}) give (\ref{z1}) for the
normalizations. \qed

\medskip
The values of the normalizations (\ref{z1}) are subject to an independent check.
Thus a well known corollary of the Selberg integral (see e.g.~\cite{Fo02,FW07p})
gives
\begin{eqnarray*}
W_{a,\beta,n} & := & \int_0^\infty dx_1 \cdots \int_0^\infty dx_n \,
\prod_{i=1}^n x_i^a e^{-x_i} \prod_{1 \le j < k \le n} |x_k - x_j|^\beta \nonumber \\
& = & \prod_{j=0}^{n-1} {\Gamma(1 + (j+1) \beta/2) \Gamma(a+1+j\beta/2) \over
\Gamma(1 + \beta/2) }.
\end{eqnarray*}
On the other hand, a simple change of variables in the definitions of $C_{\beta,n}$
as implied by Theorem \ref{one}, remembering too that the eigenvalues
therein are assumed ordered, shows
$$
C_{\beta,2m} = {1 \over 2^m m!} W_{\beta/4 - 1,\beta,m} \qquad
C_{\beta, 2m+1} = {1 \over 2^m m!} W_{3 \beta/4 -1,\beta,m}.
$$
This is consistent with (\ref{z1}).

\section{Method I part (ii): A random corank 1 projection}\label{peter_proof2}
As shown in \cite{FR02b}, in relation to the construction analogous to (\ref{Ak})
in the case of the Gaussian $\beta$-ensemble, it is possible to deduce a three term
recursion for the characteristic polynomial $P_n(x)$ and thus associate the
inductive construction with a random tridiagonal matrix. The key idea for this purpose is
to apply what may be regarded as the inverse operation of bordering, namely a random
corank 1 projection
\begin{equation}\label{pw}
\Pi_{n+1} {\rm diag} \, A_{n+1} \Pi_{n+1}, \qquad \Pi_{n+1} = {\mathbb I}_{n+1} -
{\mathbf{u} \mathbf{u}^T \over \| \mathbf{u} \|^2},
\end{equation}
for a suitable random vector $\mathbf{u}$.

The projected matrix will always have a zero eigenvalue. We want the remaining
eigenvalues to be identical in distribution to ${\rm diag} \, A_n$. To determine
the necessary form of $\mathbf{u}$ we note that we must have 
$$
\left ( {\mathbb I}_{n+1} - \left [
\begin{array}{ll} \mathbb O_n & \mathbf{0} \\
 \mathbf{0}^T & 1 \end{array} \right ] \right ) A_{n+1}
\left ( {\mathbb I}_{n+1} - \left [
\begin{array}{ll} \mathbb O_n & \mathbf{0} \\
 \mathbf{0}^T & 1 \end{array} \right ] \right ) =
\left [ \begin{array}{ll} A_n & \mathbf{0} \\
 \mathbf{0}^T & 0 \end{array} \right ].
$$
Next we write
$$
A_{n+1} = V {\rm diag} \, A_{n+1} V^\dagger
$$
where $V$ is the $(n+1) \times (n+1)$ unitary matrix of eigenvectors. We see
thus immediately that we can choose $\mathbf{u}^T = - i \mathbf{v}^T$, where $\mathbf{v}^T$ is the final row in $V$, and thus the vector of final components of the eigenvectors of $A_{n+1}$. The
structures (\ref{ww}) and (\ref{3.5}) tell us that the latter can be chosen
to have the form 
\begin{equation}\label{pw1}
\left \{
\begin{array}{ll} (iq_1 , - iq_1 ,\dots, i q_{(n+1)/2} , -iq_{(n+1)/2} ), & (n+1) \: \: {\rm even} \\
(iq_1,-iq_1,\dots,iq_{n/2},-iq_{n/2},ic), & (n+1) \: \: {\rm odd} \end{array} \right.
\end{equation}
where $q_i > 0$, $c>0$, normalized so that
\begin{equation}\label{pw2}
2 \sum_{j=1}^{(n+1)/2} q_i^2 = 1, \qquad 2 \sum_{j=1}^{n/2} q_i^2 + c^2 = 1~,
\end{equation}
depending on whether $(n+1)$ is even or odd.

Substituting $-i$ times (\ref{pw1}) for $\mathbf{u}^T$ in (\ref{pw1}), the result of
\cite[Lemma 1]{FR02b} for the characteristic polynomials of the original matrix $A_{n+1}$,
and the projected matrix, which by the choice (\ref{pw1}) 
has the same eigenvalues as $A_n$, shows
\begin{equation}\label{pcp}
{P_n(\lambda) \over \lambda P_{n+1}(\lambda) } =
\left \{
\begin{array}{ll} \displaystyle {c^2 \over \lambda^2} +
\sum_{i=1}^{n/2} {2 q_i^2  \over \lambda^2 - \lambda_i^2}, &  (n+1) \: \: {\rm odd} \\
\displaystyle
\sum_{i=1}^{(n+1)/2} {2 q_i^2  \over \lambda^2 - \lambda_i^2}, &  (n+1) \: \: {\rm even}
\end{array} \right.
\end{equation}
where $\{\lambda_i\}$ are the positive eigenvalues of $A_{n+1}$. Rather than compute the
distribution of $\{q_i^2\}$ and $c^2$ directly, we will make a trial choice and show
that it leads to the correct joint distribution of the zeros of $P_n(\lambda)$ and
$P_{n+1}(\lambda)$, which through (\ref{pcp}) uniquely determines the distribution
of $\{q_i^2\}$ and
$c^2$.

Our trial choice is to have each $2q_i^2$ distributed as $2 w_i^2$ in (\ref{w1}), (\ref{w2}),
and to have $c^2$ distributed as $b^2$ in (\ref{w2}), but with the further constraint of
the normalization conditions (\ref{pw2}). Since a normalized multivariate gamma
distribution is a Dirichlet distribution $D_n[s_1,\dots,s_n]$, our trial choice is that
for $(n+1)$ odd $\{2q_i^2\}_{i=1,\dots,n/2}\} \cup \{c^2\}$ is distributed 
according to $D_{n/2 + 1}[(\beta/2)^{n/2},\beta/4]$ (here 
$(\beta/2)^{n/2}$ is shorthand for
$\beta/2$ repeated $n/2$ times), while for $(n+1)$ even 
$\{2q_i^2\}_{i=1,\dots,(n+1)/2}$
is  distributed according to $D_{(n+1)/2}[(\beta/2)^{(n+1)/2}]$. In this
circumstance, the conditional PDF of the positive zeros of $P_n(\lambda)$ (and thus
the positive eigenvalues of $A_n$) given the zeros of $P_{n+1}(\lambda)$, or equivalently
the zeros of the random rational function in (\ref{pcp}), can be obtained by appealing to
a known result, implicit in the work of Dixon \cite{Di05} and Anderson \cite{An91},
and given explicitly in \cite[sentence below (4.10)]{FR02b}.

\begin{prop}\label{prr}
Consider the random rational function
$$
\tilde{R}(\lambda) = \sum_{i=1}^n {c_i \over \lambda - a_i}
$$
where each $a_i$ is real and $\{c_i\}$ has the Dirichlet distribution
$D_n[s_1,\dots,s_n]$. This function has exactly $(n-1)$ roots, each of 
which
is real, and for given $\{a_i\}$ these roots have the PDF
\begin{equation}
{\Gamma(s_1 + \cdots + s_n) \over \Gamma(s_1) \cdots \Gamma(s_n) }
{\prod\limits_{{\tiny 1 \le i < j \le n - 1}} (x_i - x_j) \over \prod\limits_{{\tiny 1 \le i < j \le n}}
(a_i - a_j)^{s_i + s_j - 1} }
\prod_{i=1}^{n-1} \prod_{j=1}^n | x_i - a_j|^{s_j - 1}
\end{equation}
supported on
\begin{equation}\label{4.3a}
a_1 > x_1 > a_2 > x_2 \cdots > x_{n-1} > a_n.
\end{equation}
\end{prop}
For $(n+1)$ odd, Proposition \ref{prr} applies to (\ref{pcp}) with 
$\lambda \mapsto \lambda^2$,
$a_i \mapsto \lambda_i^2$ $(i=1,\dots,n/2)$, $a_{n/2+1} = 0$, and $s_i = \beta/2$
$(i=1,\dots,n/2)$, $s_{n/2 + 1} = \beta/4$. For $(n+1)$ even, for 
application to (\ref{pcp})
we require $\lambda \mapsto \lambda^2$, $a_i \mapsto \lambda_i^2$ $(i=1,\dots,(n+1)/2)$
and $\{c_i\}_{i=1,\dots,(n+1)/2}$ distributed as $s_i = \beta/2$
$(i=1,\dots,(n+1)/2)$. We can now read off the sought conditional PDFs.

\begin{prop}\label{po}
Assume the validity of our trial choice of distribution for $2 q_i^2$ and $c^2$.
For $(n+1)$ odd, the
PDF of the positive eigenvalues of $A_n$, given that the positive eigenvalues of $A_{n+1}$
are $\lambda_1,\dots\lambda_{n/2}$, is equal to
\begin{equation}\label{m1}
{2^{n/2} \Gamma((n+1)\beta/4) \over (\Gamma(\beta/2))^{(n-1)/2} \Gamma(\beta/4) }
\prod_{i=1}^{n/2} {x_i^{\beta/2-1} \over \lambda_i^{2(3\beta/4-1)}}
\prod_{1 \le i < j \le n/2} {(x_i^2 - x_j^2) \over (\lambda_i^2 - \lambda_j^2)^{\beta - 1} }
\prod_{i,j=1}^{n/2} | x_i^2 - \lambda_j^2|^{\beta/2 - 1}
\end{equation}
where
\begin{equation}\label{m2}
\lambda_1 > x_1 > \lambda_2 > \cdots \lambda_{n/2} > x_{n/2} > 0.
\end{equation}
For $(n+1)$ even, the PDF of the positive eigenvalues of $A_n$, given that 
the positive
eigenvalues of $A_{n+1}$ are $\lambda_1,\dots,\lambda_{(n+1)/2}$, is
equal to
\begin{equation}\label{m3}
{2^{(n+1)/2} \Gamma((n+1)\beta/4) \over (\Gamma(\beta/2))^{(n+1)/2}  }
\prod_{i=1}^{(n-1)/2} x_i
{\prod\limits_{{\tiny 1 \le i < j \le (n-1)/2}} (x_i^2 - x_j^2) \over
\prod\limits_{{\tiny 1 \le i < j \le (n+1)/2}} (\lambda_i^2 - \lambda_j^2)^{\beta - 1} }
\prod_{i=1}^{(n-1)/2} \prod_{j=1}^{(n+1)/2} | x_i^2 - \lambda_j^2|^{\beta/2 - 1}
\end{equation}
where
\begin{equation}\label{m4}
\lambda_1 > x_1 > \lambda_2 > \cdots \lambda_{(n-1)/2} > x_{(n+1)/2} > 0.
\end{equation}
\end{prop}

Multiplying (\ref{m1}) and (\ref{m2}) by (\ref{pn1}) $(n \mapsto n+1)$ and
(\ref{pn2})  $(n \mapsto n+1)$ respectively gives the joint PDFs
\begin{equation}
 {1 \over \tilde{C}_{n+1,\beta} } \prod_{i=1}^{n/2} e^{- \lambda_i^2} \lambda_i
x_i^{\beta/2 - 1}
\prod_{1 \le i < j \le n/2} (\lambda_i^2 - \lambda_j^2)
(x_i^2 - x_j^2) \prod_{i,j=1}^{n/2} |x_i^2 - \lambda_j^2|^{\beta/2 - 1},
\end{equation}
$n$ even, subject to the interlacing (\ref{m2}), and
\begin{eqnarray}
&& {1 \over \tilde{C}_{n+1,\beta} } \prod_{i=1}^{(n+1)/2} e^{- \lambda_i^2} \lambda_i^{\beta/2-1}
\prod_{k=1}^{(n-1)/2} x_{k} \nonumber \\
&& \qquad \times
\prod_{1 \le i < j \le (n+1)/2} (\lambda_i^2 - \lambda_j^2)
\prod_{1 \le i' < j' \le (n-1)/2} (x_{i'}^2 - x_{j'}^2)
\prod_{i,j=1}^{n/2} |x_i^2 - \lambda_j^2|^{\beta/2 - 1},
\end{eqnarray}
$n$ odd, subject to the interlacings (\ref{m4}). Here
\begin{eqnarray*}
&& \tilde{C}_{n+1,\beta} =
C_{n+1,\beta} {(\Gamma(\beta/2))^{(n-1)/2} \Gamma(\beta/4) \over
2^{n/2} \Gamma((n+1)\beta/4) },  \qquad (n+1) \quad {\rm odd} \\
&& \tilde{C}_{n+1,\beta} =
C_{n+1,\beta} {(\Gamma(\beta/2))^{(n+1)/2}  \over
2^{(n+1)/2} \Gamma((n+1)\beta/4) },  \qquad (n+1) \quad {\rm even.}
\end{eqnarray*}
On the other had the same joint PDFs can be obtained by interchanging the
symbols $x_i \leftrightarrow \lambda_i$ in (\ref{x1}) and (\ref{x2}), and
multiplying by (\ref{pn1}) (with $\lambda_j \mapsto x_j$) and
(\ref{pn2}) (with $\lambda_j \mapsto x_j$). This tells us that our trial
choice of the distributions of the components of the eigenvectors is
correct, and so in particular the qualification starting off the statement
of Proposition \ref{po} can be removed.

The fact that our trial choice of the distributions of the components of the eigenvectors
is correct also implies that we can make the replacements in (\ref{pcp})  as
indicated in the first sentence of the paragraph below (\ref{pcp}), and so obtain
\begin{equation}\label{npn}
{\mathcal N}_n {P_n(\lambda) \over \lambda P_{n+1}(\lambda) } =
\left \{
\begin{array}{ll} \displaystyle
 {b^2 \over \lambda^2} +
\sum_{i=1}^{n/2} {2 w_i^2  \over \lambda^2 - \lambda_i^2}, &  (n+1) \: \: {\rm odd} \\
\displaystyle
\sum_{i=1}^{(n+1)/2} {2 w_i^2  \over \lambda^2 - \lambda_i^2}, &  (n+1) \: \: {\rm even}
\end{array} \right.
\end{equation}
where
\begin{equation}\label{npn1}
{\mathcal N}_n =
\left \{
\begin{array}{ll} \displaystyle
b^2  +
\sum_{i=1}^{n/2} 2 w_i^2  , &  (n+1) \: \: {\rm odd} \\
\displaystyle
\sum_{i=1}^{(n+1)/2} 2 w_i^2 , &  (n+1) \: \: {\rm even}
\end{array} \right.
\end{equation}
We can substitute (\ref{npn}) with $n \mapsto (n - 1)$ in
(\ref{w1}), (\ref{w2}) to deduce a random three term recurrence for $\{P_n(\lambda)\}$.

\begin{prop}\label{p4.3}
The characteristic polynomials $\{P_n(\lambda)\}_{n=2,3,\dots}$ for the matrices
$\{A_n\}_{n=2,3,\dots}$ defined inductively in Proposition \ref{p3.1} satisfy
the  random three term recurrence
\begin{equation}\label{pbp}
P_{n+1}(\lambda) = \lambda P_n(\lambda) - b_{n}^2 P_{n-1}(\lambda), \qquad
n=1,2,\dots
\end{equation}
with $P_0(\lambda) := 1$, $P_1(\lambda) := \lambda$ and where $b_n$ has
distribution $\Gamma[n\beta,1]$. This is the three term recurrence for the
characteristic polynomial of the tridiagonal matrix (\ref{3.3}).
\end{prop}

\noindent
{\it Proof.} \quad The substitution gives
$$
{P_{n+1}(\lambda) \over \lambda P_n(\lambda) } = 1 -
{\mathcal N}_{n-1} {P_{n-1}(\lambda) \over \lambda P_n(\lambda) }
$$
and so we can obtain (\ref{pbp}) with $b_{n}^2 = {\mathcal N}_{n}$. The
distribution of $b_{n}^2$ then follows from (\ref{npn1}) and the fact that
the number of degrees of freedom in the sum of independent gamma distributed
variables adds. That the three term recurrence relates to an
anti-symmetric tridiagonal matrix is a standard result. \qed

Combining Proposition \ref{p4.3} with (\ref{pcp}) allows the distribution of the
first element in each of the independent eigenvectors of (\ref{3.3}) to be determined. We can thus show that the vector of first components of the independent eigenvectors of the random tridiagonal matrix
(\ref{3.3}), which we choose to be positive, has the Dirichlet distribution given by Theorem \ref{two}.

\vspace{.25cm}

\noindent
\underline{{\it {\bf First proof of Theorem \ref{two}.} }} For a general $n \times n$ real symmetric matrix $X$ we have that
\begin{equation}\label{5.5}
{P_{n-1}(\lambda) \over P_n(\lambda)} = \sum_{i=1}^n {c_i \over \lambda - \lambda_i}
\end{equation}
where $P_n(\lambda)$ is the characteristic polynomial of $X$, $P_{n-1}(\lambda)$ is the
characteristic polynomial of the submatrix formed from $X$ by blocking the first row and
first column, $\{\lambda_i\}$ are the eigenvalues of $X$, and $\{c_i\}$ are
the first component of the eigenvectors. For the matrix (\ref{3.3}), the eigenvalues
and eigenvectors have the special structures (\ref{3.5}) and (\ref{ww})
respectively. Substituting in (\ref{5.5}) we reclaim again (\ref{pcp}) provided
the independent members of $\{c_i\}$ are identified with the independent entries in
(\ref{pw1}) (which are the last components of the eigenvectors of (\ref{Ak})).
But we know from the paragraph below Proposition \ref{po} that the latter entries
have the Dirichlet distributions as claimed. \qed

\section{Method II: Jacobians of anti-symmetric tridiagonal matrices} \label{ioana_proof1}

In this section, we present an alternative proof of Theorems \ref{one} 
and \ref{two}, based on the mapping between a real 
anti-symmetric tridiagonal matrix and its positive eigenvalues and first row of the eigenvector matrix. This proof 
is very much in the spirit of \cite{DE02}; in fact, many of the results used there for symmetric matrices can be 
used here for anti-symmetric ones, with very minor modifications. In the interest of brevity, we present the 
proofs only if the modifications are non-trivial.

We start by giving a set of linear algebra results, which build up to the computation of the eigenvalue PDF for the 
random tridiagonal matrices \eqref{3.3}.

Anti-symmetric matrices are \emph{normal} matrices, i.e., they have the property that they commute with their transpose (for a treatment of normal matrices, see any linear algebra book, e.g. \cite{GVL96}). Equivalently, they have the very special property of being diagonalizable via a unitary transformation. Any real anti-symmetric matrix $T$ can be decomposed as
$T = U \Lambda U^{H}$, with $U$ a unitary matrix and $\Lambda$ the diagonal matrix of eigenvalues.

Let $T$ be a real anti-symmetric tridiagonal matrix in \emph{reduced} form, defined as
\begin{eqnarray} \label{def_trid}
T & = & \left [ \begin{array}{ccccc} 0 & b_{n-1} & & & \\ -b_{n-1} & 0 &  b_{n-2} & & \\ & \ddots & \ddots & \ddots & \\ & & -b_2 & 0 & b_1 \\ & & & -b_1 & 0 \end{array} \right ]~;
\end{eqnarray}
if $b_i>0$ for all $i=1,\dots,n$, then when $n$ is even $T$ is full-rank, whereas when $n$ is odd $0$ is a simple eigenvalue.

As mentioned already in the introduction, the non-zero eigenvalues for such matrices come in pairs $(i 
\lambda_j, -i \lambda_j)$, with $j =1,\dots,\lr \frac{n}{2} \rr$, and 
we assume the ordering $\lambda_1 > \lambda_2 > 
\ldots > \lambda_{\lr \frac{n}{2} \rr} > 0$.

The decomposition $T = U \Lambda U^{H}$ is not unique; to make it unique, we impose two conditions. First, 
we order 
the diagonal of $\Lambda$ as follows: $(i\lambda_1, \ldots, i \lambda_{\lr \frac{n}{2} \rr}, -i\lambda_1, 
\ldots, -i\lambda_{\lr \frac{n}{2} \rr})$. 
If $n$ is odd, we let $\Lambda(n,n) = 0$. (Note that this ordering is different to that used in
Proposition \ref{p3.1}, but is more convenient for present purposes.)

Second, we impose the condition that the first row of $U$ has positive entries (note that eigenvectors are only defined up to 
multiplication by rotations $e^{i \theta}$, and also that if $u_j$ is an eigenvector for eigenvalue $i \lambda_j$, 
then $\overline{u_j}$, the conjugate of $u_j$, is an eigenvector for $-i \lambda$).

The first result we state is very similar to Theorem 7.2.1 in \cite{Pa98}, and the proof given in \cite{Pa98} is adaptable almost verbatim to this case (thus we choose not to repeat it).

\begin{lemma} \label{ext_map}
With the notation above, let $q:=(q_1, \ldots, q_n)$ be the first row of the matrix $U$, and $\lambda$ be the diagonal of $-i\Lambda$. Then $T$ is uniquely determined by $\lambda$ and $q$.
\end{lemma}

Due to the real anti-symmetric nature of $T$, we can in fact deduce that

\begin{cor} \label{short_map}
$T$ is determined by $\lambda_1, \ldots, \lambda_{\lr \frac{n}{2} \rr}$, $q_1, \ldots, q_{\lr \frac{n}{2} \rr}$. 
\end{cor}

\begin{proof} 
Note that by our choice of $U$ and $\Lambda$, $q_{\lr \frac{n}{2} \rr +j} 
= q_j$ for all $j = 1,\dots,\lr \frac{n}{2} \rr$. The statement is 
immediate for $n$ even; if $n$ is odd, note that $0$ is also an eigenvalue 
and that the first component $q_n$ corresponding to it can be determined 
from the condition $\sum\limits_{j=1}^n q_j^2 = 1$. 
\end{proof}

Before we proceed, we will need the following lemma, which is similar to Lemma 2.7 of \cite{DE02}. 

\begin{lemma} \label{vander}
Let $\Delta(\lambda^2) := \prod_{i<j} (\lambda_i^2 - \lambda_j^2)$, where $\lambda_1, \ldots, \lambda_{\lr \frac{n}{2} \rr}$ are the positive imaginary parts of the eigenvalues of $T$, and $q$ is as above. Then \begin{itemize} \item if $n=2k$, 
\[
(\Delta(\lambda^2))^2 = \frac{\prod\limits_{i=1}^{n-1} b_i^{i}}{2^{k} \prod\limits_{i=1}^k \left (q_i^2 \lambda_i \right)}~~.
\]
\item if $n=2k+1$,
\[
(\Delta(\lambda^2))^2 = \frac{\prod\limits_{i=1}^{n-1} b_i^{i}}{2^{k} \cdot q_n \cdot \prod\limits_{i=1}^k \left (q_i^2 \lambda_i^3 \right)}~~.
\]
\end{itemize}
\end{lemma}

\begin{proof} To keep eigenvalues real, we will examine the matrix $iT$. Just as in \cite{DE02}, we use the three-term recurrence for the characteristic polynomial of $iT$.  Denote by $\lambda_i^{(m)},~i=1,\dots,m$ the eigenvalues of the $m \times m$ lower corner submatrix of $iT$, and denote by 
$P_m(x) = \prod\limits_{i=1}^m (x-\lambda_i^{(m)})$ the corresponding characteristic polynomial. Then the
three term recurrence (\ref{pbp}) holds,
and from this, writing 
\begin{eqnarray} \label{doi}
\prod_{1 \leq i \leq m \atop 1 \leq j \leq m-1} 
|\lambda_i^{(m)} - \lambda_{j}^{(m-1)}| = \prod_{i=1}^m |P_{m-1}(\lambda_i^{m})| = \prod_{j=1}^{m-1} |P_m(\lambda_j^{(m-1)}|~,
\end{eqnarray} 
we deduce that 
\begin{eqnarray} \label{trei}
\left | \prod_{i=1}^{m-1} P_m(\lambda_i^{(m-1)}) \right| = b_{m-1}^{2(m-1)} ~\cdot~\left | \prod_{i=1}^{m-1} P_{m-2}(\lambda_i^{(m-1)}) \right|~.
\end{eqnarray}

By repeatedly applying \eqref{pbp} and \eqref{doi}, we obtain that 
\begin{eqnarray} \label{patru}
\prod_{i=1}^{n-1} |P_n(\lambda_i^{(n-1)})| = \prod_{i=1}^{n-1} b_i^{2i}~.
\end{eqnarray}

Like in \cite{DE02}, we make use of a simple identity for $q_i^2$, which is a particular form of Theorem 7.9.2 from \cite{Pa98}:
\begin{eqnarray} \label{cinci}
q_i^2 = \left | \frac{P_{n-1}(\lambda_i^{(n)})}{P'_n(\lambda_i^{(n)})} \right |~~, ~~\forall~1 \leq i \leq n~.
\end{eqnarray}

Note that \eqref{cinci} can also be seen as a corrolary of \eqref{pcp}.

Let us now examine $P'_n(\lambda_i)$, with $\lambda_i = \lambda_i^{(n)}$. Since
\[P'_n(x) = \sum_{i=1}^n \prod_{j \neq i} (x - \lambda_i)~~,
\]
it follows that \begin{itemize} \item if $n=2k$, $|P'_n(\lambda_i)| = 2|\lambda_i| 
\prod\limits_{{j=1\atop j \neq i }}^{k} |\lambda_i^2 - \lambda_j^2|$, for all $i= 
1,\dots,k$, and
\item if $n = 2k+1$, \begin{itemize} \item[$\circ$] $|P'_n(\lambda_i)| = 2 \lambda_i^2 
\prod\limits_{{j=1\atop j \neq i }}^{k} |\lambda_i^2 - \lambda_j^2|$, for all $i= 
1,\dots,k$, and
 \item[$\circ$] $|P'_n(0)| = \prod\limits_{i=1}^{k} \lambda_i^2$.
\end{itemize}
\end{itemize}

Since $q_{\lr \frac{n}{2} \rr+j} = q_j$, for all $j = 1,\dots,\lr \frac{n}{2} \rr$, it follows that \begin{itemize} \item if $n = 2k$, 
\begin{eqnarray} \label{sase}
\prod_{i=1}^{k} q_i^4 =  \frac{\prod\limits_{i=1}^{n} |P_{n-1}(\lambda_i)|}{2^n (\Delta(\lambda^2))^4 \prod\limits_{i=1}^k \lambda_i^2} = \frac{\prod\limits_{i=1}^n b_i^{2i}}{2^n (\Delta(\lambda^2))^4 \prod\limits_{i=1}^k \lambda_i^2}~~,
\end{eqnarray}
\item if $n = 2k+1$, 
\begin{eqnarray} \label{sapte}
q_n^2 \cdot \prod_{i=1}^{k} q_i^4 =  \frac{\prod\limits_{i=1}^{n} |P_{n-1}(\lambda_i)|}{2^{n-1} 
(\Delta(\lambda^2))^4 \prod\limits_{i=1}^k \lambda_i^6} =  \frac{\prod\limits_{i=1}^n b_i^{2i}}{2^n (\Delta(\lambda^2))^4 \prod\limits_{i=1}^k \lambda_i^6}~~.
\end{eqnarray}
\end{itemize}

Rewriting \eqref{sase} and \eqref{sapte} and taking square roots, one obtains the statement of the lemma. \end{proof}

Consider now the transformation $T \leftrightarrow (q, \lambda)$, with all of the conditions imposed above; the transformation is one-to-one and onto, and it must thus have a Jacobian $\mathcal{J}$. We will compute this Jacobian in the same way we computed the Jacobian of the similar transformation for symmetric tridiagonal matrices in \cite{DE02}: we will make use of the fact established in
Section \ref{house_red} that the anti-symmetric GUE ensemble has the same eigenvalue distribution as the tridiagonal model \eqref{3.3} with $\beta = 2$. 

An alternative derivation for the Jacobian $\mathcal{J}$ is given in the 
Appendix.

Specifically, we will require the following three properties of the anti-symmetric GUE ensemble $(\beta = 2)$.

\begin{itemize}
\item[\emph{Property 1.}] The eigenvalue distribution of the anti-symmetric GUE ensemble is given by \eqref{3.3} with $\beta = 2$.
\item[\emph{Property 2.}] The set of eigenvalues and the set of eigenvectors of the anti-symmetric GUE ensemble are 
statistically independent of each other. 
\item[\emph{Property 3.}] 

$\bullet$ For $n=2k$, let 
$U = [{\mathbf u}_1, {\mathbf u}_2, \ldots, {\mathbf u}_k, \overline{{\mathbf u}_1}, 
\overline{{\mathbf u}_2}, \ldots, \overline{{\mathbf u}_k}]$ be a matrix of unit-norm eigenvectors. For each 
$j =1,\dots,k$, write ${\mathbf u}_j = {\mathbf v}_j + i {\mathbf w}_j$, 
with ${\mathbf v}_j$ and ${\mathbf w}_j$ real. Then, with probability $1$, 
the set $\{{\mathbf v}_j, j = 1,\dots,k\} \cup \{{\mathbf w}_j, j = 1,\dots,k\}$ is an orthogonal basis for $O(2k)$. Moreover, 
for any $j$, $\sqrt{2}{\mathbf v}_j$ and $\sqrt{2}{\mathbf w}_j$ are distributed uniformly over the sphere.

\vspace{.15cm}

\noindent $\bullet$ For $n=2k+1$, let $U = [{\mathbf u}_1, {\mathbf u}_2, \ldots, 
{\mathbf u}_k, \overline{{\mathbf u}_1}, \overline{{\mathbf u}_2}, \ldots, \overline{{\mathbf u}_k}, z]$ be 
a matrix of unit-norm eigenvectors. For each $j =1,\dots,k$, write ${\mathbf u}_j = \mathbf{v}_j 
+ i {\mathbf w}_j$, with ${\mathbf v}_j$ and ${\mathbf w}_j$ real. 
Then, with probability $1$, the set $\{{\mathbf v}_j, j = 1,\dots,k\} \cup \{{\mathbf w}_j, j = 1,\dots,k\} 
\cup \{z\}$ is an orthogonal basis for $O(2k+1)$. Moreover, for any $j$, $\sqrt{2}{\mathbf v}_j$ and 
$\sqrt{2}{\mathbf w}_j$ are 
distributed uniformly over the sphere, and $z$ is distributed uniformly over the sphere.

\end{itemize}

Choose now $U$ to be the (unique!) matrix of anti-symmetric GUE eigenvectors which has on its first row all positive 
numbers $q_1, \ldots, q_n$. From the three properties above one can immediately deduce Proposition \ref{distr}.

\begin{prop} \label{distr}
If $n=2k$, $(q_1, \ldots, q_n)$ has the same distribution as 
$\frac{({\mathbf w},
{\mathbf w})}{||({\mathbf w}, {\mathbf w})||_2}$, where $\mathbf w$ is a 
vector of $k$ independent variables with distribution ${\mathbf w} \sim (\chi_2, \chi_2, \ldots, \chi_2)$.

If $n = 2k+1$, $(q_1, \ldots, q_n)$ has the same distribution as 
$\frac{({\mathbf w}, {\mathbf w}, z)}{||({\mathbf w},
{\mathbf w},z)||_2}$, where $\mathbf w$  is a vector of $k$ independent variables with 
distribution ${\mathbf w} \sim (\chi_2, \chi_2, \ldots, \chi_2)$, and $z$ is a scalar $\chi_1$-distributed variable independent of $w$. 
\end{prop}

We now proceed to compute the Jacobian of the transformation $T \leftrightarrow (q, \lambda)$.

The joint element density of a matrix $T$ from the distribution \eqref{3.3} with $\beta = 2$ is
\begin{eqnarray} \label{zero}
\mu(b) ~~=~ \frac{2^{n-1}}{\prod\limits_{i=1}^{n-1} \Gamma \left (\frac{i}{2} \right )}~\prod_{i=1}^{n-1} b_i^{i-1} e^{-\sum\limits_{i=1}^{n-1} b_i^2}~.
\end{eqnarray}

Denote by d$b = \wedge_{i=1}^{n-1}$d$b_i$, d$\lambda = \wedge_{i=1}^{\lr \frac{n}{2} \rr} \lambda_i$. To be consistent with Property 3, denote by 
${\rm d}q$
\begin{itemize} \item if $n=2k$, $dq$ is the surface element on the $k$-dimensional sphere of radius $1/2$,
\item if $n=2k+1$, ${\rm d}q$ is the 
surface element on the $(k+1)$-dimensional ellipsoid of first $k$ axes equal to $1/2$, and the $(k+1)$st equal to $1$.
\end{itemize}

With the transformation $T \leftrightarrow (q, \lambda)$, it follows that
\begin{eqnarray} \label{einz}
\mu(b)~db = \mathcal{J}~\mu(b(q, \lambda))~\mbox{d}q \mbox{d}\lambda \equiv \nu(q, \lambda)~\mbox{d}q \mbox{d}\lambda~~.
\end{eqnarray}

By Proposition \ref{distr} and Properties 1, 2, and 3, it follows that 
\begin{itemize}
\item if $n = 2k$, 
\begin{eqnarray} \label{zwei}
\nu(q, \lambda) = \frac{1}{C_{2, 2k}} ~\left(\Delta(\lambda^2)\right)^2 ~~e^{-\sum\limits_{i=1}^k 
\lambda_i^2} ~~~~\frac{\Gamma(k)}{2} ~\prod\limits_{i=1}^k q_i~;
\end{eqnarray}
\item if $n = 2k+1$, 
\begin{eqnarray} \label{drei}
\nu(q, \lambda) = \frac{1}{C_{2, 2k+1}} ~\left(\Delta(\lambda^2)\right)^2 ~\prod\limits_{i=1}^k 
\lambda_i^2~~e^{-\sum\limits_{i=1}^k \lambda_i^2}~~~~\frac{ \Gamma \left(\frac{2k+1}{2} 
\right)}{\Gamma \left( \frac{1}{2} \right)}~ \prod\limits_{i=1}^k q_i~. \end{eqnarray}
\end{itemize}

Hence all that is left is to compute the Jacobian $\mathcal{J}$ as
\[
\mathcal{J} = \frac{\nu(q,\lambda)}{\mu(b)}~.
\]

Since the Frobenius norm of a matrix is preserved by orthogonal similarity transformations, 
\begin{eqnarray} \label{vier}
 \sum_{i=1}^n b_i^2 = \frac{1}{2} ||T||_F = \frac{1}{2} ||\Lambda||_F = \sum_{i=1}^{\lr \frac{n}{2} \rr} \lambda_i^2~.
\end{eqnarray}

By putting together \eqref{zero}-\eqref{vier}, Lemma \ref{vander}, and the definition of $C_{2, n}$ from Theorem \ref{one}, all constants cancel and we obtain the following lemma.

\begin{lemma} \label{jacobian}
The Jacobian $\mathcal{J}$ of the transformation $T \leftrightarrow (q, \lambda)$ is given by \begin{itemize}
\item if $n = 2k$, 
\[
\mathcal{J} = \frac{\prod\limits_{i=1}^{n-1} b_i}{\prod\limits_{i=1}^k q_i \lambda_i}~,
\]
\item if $n = 2k+1$,
\[
\mathcal{J} = \frac{\prod\limits_{i=1}^{n-1} b_i}{q_n \prod\limits_{i=1}^k q_i \lambda_i}~.
\]\end{itemize}
\end{lemma}

We are now able to give an alternate proof of Theorems \ref{one} and \ref{two}.

\vspace{.25cm}

\noindent \underline{{\bf \textit{Second proof of Theorems \ref{one} and \ref{two}}}}. Starting from the joint element density of the matrix $A_{n}^{\beta}$, 
\[
\mu_{n, \beta} (b) = \frac{2^{n-1}}{\prod\limits_{i=1}^{n-1} \Gamma \left( \frac{i \beta}{2} \right)} ~\prod\limits_{i=1}^{n-1} b_i^{\frac{i \beta}{2}-1} ~~e^{ - \sum\limits_{i=1}^{n-1} b_i^2}~,
\]
we make the transformation $A_{n}^{\beta} \leftrightarrow (q, \lambda)$, into the eigenvalues and first row of the eigenvector matrix; we use the computed Jacobian from Lemma \ref{jacobian} and the expression for the Vandermonde from Lemma \ref{vander}. 

\begin{itemize} \item For $n = 2k$, we obtain the joint density of $q$ and $\lambda$:
\begin{eqnarray*}
\nu_{n, \beta}(q, \lambda) & =  & \mathcal{J} \mu_{n, \beta}(b(q, \lambda)) \\
& = & \frac{2^{n-1}}{\prod\limits_{i=1}^{n-1} \Gamma \left( \frac{i \beta}{2} \right)} ~
\frac{\prod\limits_{i=1}^{n-1} b_i^{\frac{i \beta}{2}}}{ \prod\limits_{i=1}^{k} q_i \lambda_i} ~~e^{- \sum\limits_{i=1}^{n-1} \lambda_i^2}\\
& = & \frac{2^{n-1}}{\prod\limits_{i=1}^{n-1} \Gamma \left( \frac{i \beta}{2} \right)} ~ 
\left ( \frac{\prod\limits_{i=1}^{n-1} b_i^i}{2^k \prod\limits_{i=1}^{k} q_i^2 \lambda_i} \right)^{\beta/2} ~ 
\frac{\left (2^k \prod\limits_{i=1}^{k} q_i^2 \lambda_i \right)^{\beta/2}}{\prod\limits_{i=1}^{k} q_i \lambda_i}   ~~e^{- \sum\limits_{i=1}^{n-1} \lambda_i^2}~ \\
& = & \frac{2^{n-1 + k \beta/2}}{\prod\limits_{i=1}^{n-1} \Gamma \left( \frac{i \beta}{2} \right)} ~
\left ( \left(\Delta(\lambda^2) \right)^{\beta}~\prod\limits_{i=1}^{k} 
\lambda_i^{\beta/2-1} ~~e^{- \sum\limits_{i=1}^{n-1} \lambda_i^2} 
\right ) ~~~ \left (\prod\limits_{i=1}^{k} q_i^{\beta-1} \right )~.
\end{eqnarray*}

It is easy to check that 
\[
 \frac{2^{n-1 + k \beta/2}}{\prod\limits_{i=1}^{n-1} ~\Gamma \left( \frac{i \beta}{2} \right)}= 
\frac{1}{C_{\beta, 2k}} \frac{2^{k-1+k\beta/2} \Gamma \left ( \frac{k \beta}{2} \right)}{\left ( \Gamma \left ( \frac{\beta}{2} \right) \right)^{k}}~;
\]
this shows that $\lambda$ and $q$ decouple and that they have the distributions described in Theorem \ref{one}
and \ref{two}.

\item For $n=2k+1$, the joint density on $q$ and $\lambda$ can be obtained as
\begin{eqnarray*}
\nu_{n, \beta}(q, \lambda) & =  & \mathcal{J} \mu_{n, \beta}(b(q, \lambda)) \\
& = & \frac{2^{n-1}}{\prod\limits_{i=1}^{n-1} \Gamma \left( \frac{i \beta}{2} \right)} ~
\frac{\prod\limits_{i=1}^{n-1} b_i^{\frac{i \beta}{2}}}{q_n \prod\limits_{i=1}^{k} q_i \lambda_i} ~~e^{-\sum\limits_{i=1}^{n-1} \lambda_i^2}\\
& = & \frac{2^{n-1}}{\prod\limits_{i=1}^{n-1} \Gamma \left( \frac{i \beta}{2} \right)} ~ 
\left ( \frac{\prod\limits_{i=1}^{n-1} b_i^i}{2^k q_n \prod\limits_{i=1}^{k} q_i^2 \lambda_i^3} \right)^{\beta/2} ~ 
\frac{\left (2^k q_n \prod\limits_{i=1}^{k} q_i^2 \lambda_i^3 \right)^{\beta/2}}{q_n \prod\limits_{i=1}^{k} q_i \lambda_i}   ~~e^{- \sum\limits_{i=1}^{n-1} \lambda_i^2}~ 
\\
& = & \frac{2^{n-1+k\beta/2}}{\prod\limits_{i=1}^{n-1} \Gamma \left( \frac{i \beta}{2} \right)} ~
\left ( \left(\Delta(\lambda^2) \right)^{\beta}~\prod\limits_{i=1}^{k} 
\lambda_i^{3 \beta/2-1} ~~e^{- \sum\limits_{i=1}^{n-1} \lambda_i^2} 
\right)~~~ 
\left ( q_n^{\beta/2 - 1} \prod\limits_{i=1}^{k} q_i^{\beta-1} \right ) ~.
\end{eqnarray*}

Once can easily check that
\[
 \frac{2^{n-1}}{\prod\limits_{i=1}^{n-1} \Gamma \left( \frac{i \beta}{2} \right)}= 
\frac{1}{C_{\beta, 2k+1}} \frac{2^{k+ k \beta/2} ~\Gamma \left ( \frac{k \beta + \beta/2}{2} \right)}{\Gamma \left( \frac{\beta}{4} \right) \left ( \Gamma \left ( \frac{\beta}{2}
\right) \right)^{k}}~,
\]
so again, $\lambda$ and $q$ decouple, and they have the distributions described in 
Theorems \ref{one} and \ref{two}.
\end{itemize}

This completes the proof. \qed

\section{Method III: an orthogonal transformation} \label{ioana_proof2}

 This proof is based on the observation that the distributions of 
Theorem \ref{one} are, essentially, $\beta$-Laguerre distributions: they are the same as the 
distributions of singular values of the $B_{\beta, n, a}$, bidiagonal, root-Laguerre matrices given in \cite{DE02}. 

We describe this class of matrices below.

For $2a - (n-1)\beta > 0$, let $B_{\beta,n,a}$ be the $n \times n$ random bidiagonal matrix 
\[
B_{\beta, n, a} = \left [ \begin{array}{cccc} \chi_{2a} & & & \\
					\chi_{(n-1) \beta} & \chi_{2a-\beta} & & \\
							& \ddots & \ddots & \\
							& & \chi_{\beta} & \chi_{2a-(n-1) \beta} \end{array} \right ]~.
\]
It was proved in \cite{DE02} that the eigenvalue distribution of $L_{\beta, n, a} = B_{\beta, n, a}^{} B_{\beta, n, a}^{T}$(or, equivalently, $W_{\beta, n, a} =  B_{\beta, n, a}^{T}  B_{\beta, n, a}^{}$, since they are the squares of the singular values of $ B_{\beta, n, a}$) is the well-known Laguerre distribution of size $n$ and parameter $a-(n-1)\beta/2-1$. Denoting by $\lambda_1 > \ldots > \lambda_n >0 $ the eigenvalues of $L_{\beta, n, a}$, their joint PDF is given by
\[
f_{n, a, \beta} = c_{L}^{\beta,a} \prod_{i<j} (\lambda_i - \lambda_j)^{\beta} \prod_{i}
\lambda_i^{a - (n-1) \frac{\beta}{2}-1} ~~e^{-\sum\limits_{i=1}^n\lambda_i/2}~,~~
\]
where 
\[
c_{L}^{\beta,a} =  2^{-na} \prod_{j=1}^{n}
\frac{\Gamma \left (\frac{\beta}{2} \right )}{\Gamma \left (\frac{\beta}{2}j \right ) \Gamma \left (a - \frac{\beta}{2}(n-j)) \right )}~~.
\]

Since the singular values of $B_{\beta, n, a}$ are the square roots of the eigenvalues of $L_{\beta, n, a}$, the 
joint PDF of the singular values of $B_{\beta, n, a}$ is given by 
\begin{eqnarray} \label{svd}
\tilde{f}_{n, a, \beta} = 2^{n} c_{L}^{\beta,a} \prod_{i<j} (\sigma_i^2 - \sigma_j^2)^{\beta} \prod_{i}
\sigma_i^{2a - (n-1)\beta-1}  ~e^{-\sum\limits_{i=1}^n \sigma_i^2/2} ~.
\end{eqnarray}

The proof is based on the observation that, if we let the size of the matrix anti-symmetric matrix be $n = 2k$, with $k$ variables, the parameter $a = \frac{2k-1}{4} \beta$, and in addition scale each $\sigma_i$ by $\sqrt{2}$, the PDF of \eqref{svd} is the same as the one in Theorem \ref{one}; while, if we let $n=2k+1$ and $a = \frac{2k+1}{4} \beta$, after scaling $\sigma_i$ by $\sqrt{2}$, the PDF of \eqref{svd} is once again the same as in Theorem \ref{one}.

We first recall the following well-known result in linear algebra.

\begin{prop} \label{compsvd}
Let $Y$ be a real $n \times n$ matrix, and construct the $2n \times 2n$ matrix \[V_Y = \left [ \begin{array}{cc} 0 & -Y \\ Y^{T} & 0 \end{array} \right ]~.\] Let $\sigma_1, \ldots, \sigma_n$ be the singular values of $Y$ (with multiplicities). Then the eigenvalues of $V_Y$ are $\pm i \sigma_1, \pm i \sigma_2, \ldots, \pm i \sigma_n$ (also with multiplicities).
\end{prop}

If $Y$ is a bidiagonal matrix, one can ``shuffle'' the entries of the matrix $V_Y$ to make an (anti-symmetric) tridiagonal out of them. We first need to define ``shuffling''.

\begin{defs} \label{shuf}
We define the ``perfect shuffle'' $2n \times 2n$ permutation matrix $P_n$ to be given by 
\begin{eqnarray*}
P_n(i,j) & = & \left \{ \begin{array}{cl} 1, & \mbox{if}~j = \frac{i+1}{2}~\mbox{or}~j = n+ \frac{i}{2}~,\forall~ 2n \geq i,j \geq 1\\
		0, & \mbox{otherwise.} \end{array} \right .
\end{eqnarray*}
\end{defs}

Note that, given a matrix $X$, $P_nX$ has the same rows as $X$, but listed in the following order: $1, n+1, 2, n+2, \ldots, n, 2n$, while $XP_n^{T}$ has the same columns of $X$ but rearranged in the same order $1, n+1, 2, n+2, \ldots, n, 2n$. Also note that, since $P_n$ is a permutation, $P_n$ is orthogonal.

We can now define the \emph{alternating sign perfect shuffle} (ASPS) matrix.

\begin{defs} \label{asps} Let $D_n$ be the diagonal matrix for which $D_n(i,i) = (-1)^{\lr \frac{i}{2} \rr}$, $i = 1..2n$. We call the matrix $Q_n = D_n P_n$ the alternating sign perfect shuffle (ASPS) matrix. Note that $Q_n$ is also orthogonal.
\end{defs}

We can now explain the effect of the ASPS matrix on a tridiagonal matrix.  

\begin{lemma} \label{mult}
Let $T$ be a tridiagonal anti-symmetric matrix (as in \eqref{def_trid}). Then $T = Q_n V_{B} Q_n^{T}$, where $B$ is the $n \times n$ bidiagonal matrix 
\[
B = \left [ \begin{array}{ccccc} b_{n} & & & & \\ b_{n-1} & b_{n-2} & & & \\ & b_{n-3} & b_{n-4} & & \\ & & \ddots & \ddots & \\ & & & b_2 & b_1 \end{array} \right ]~,
\]
and $V_B$ is like in Proposition \ref{compsvd}.
\end{lemma}

The proof of Lemma \ref{mult} is an easy exercise; it suffices 
to see how the entries of $V_B$ move around under 
left multiplication by $Q_n$, respectively, right multiplication by $Q_n^{T}$.

We give here an example: for $i \leq n$, the entry $(i, i+n)$ moves first to $(2i-1, i+n)$ under the multiplication by $Q_n$ to the left, then it moves to $(2i-1, 2(i-1))$ under multiplication by $Q_n^{T}$ to the right. Along the way, it gets multiplied by $(-1)^{\lr \frac{i}{2} \rr + \lr \frac{i-1}{2} \rr}$, and thus it changes sign. 

The other cases can be examined in the same way. 

The proof for matrix ($A_n^{\beta}$) size $n = 2k$ differs slightly from the one for $n= 2k+1$; we present them separately.

\vspace{.25cm}

\noindent \underline{{\it {\bf Third proof of Theorems 
\ref{one} 
and \ref{two}, $n = 2k$.}}} Armed with Lemma \ref{mult}, Theorem \ref{one} follows directly in the case when $n= 2k$, as
\[
\sqrt{2} A_{n}^{\beta} = Q_n V_{B_{\beta, n, a}} Q_n^{T}~,
\]
where the equality should be understood in terms of distributions.

In addition, the distribution of the first components of the eigenvectors of $A_n^{\beta}$ (given by Theorem \ref{two}) can be obtained from this orthogonal similarity transformation, as a consequence of the following three facts: \begin{itemize}
\item if $B = U \Sigma V^{T}$ is the SVD of $B$, then 
\[
V_B = \left [ \begin{array}{rr} U & U \\ V & -V \end{array} \right ]  \left [ \begin{array}{rr} -\Sigma & 0 \\ 0 & \Sigma \end{array} \right ]  \left [ \begin{array}{rr} U^{T} & V^{T} \\ U^{T} & -V^{T} \end{array} \right ] ~~;
\]
is the eigenvalue decomposition for $V_B$.
\item the first row of the left singular vectors for $B_{\beta,n,a}$ is distributed like a (normalized to $1$) vector of i.i.d. $\chi_{\beta}$ random variables (see \cite{DE02});
\item the first row of the matrix $Q_n X$ is the same as for $X$, for any matrix $X$. 
\end{itemize} \qed

For the $n$ odd case, the proof is only slightly more complicated.

\noindent \underline{{\it {\bf Third proof of Theorems 
\ref{one} 
and \ref{two}, $n = 2k+1$.}}} The first obstacle in using the ASPS matrix is that the size of $A_n^{\beta}$ is odd. This is easily overcome by introducing an extra row and column of zeroes, set 
\[
\tilde{A}_{k}^{\beta} = \left [\begin{array}{cc} A_{2k+1}^{\beta} & 0_{2k} \\
					       0_{2k}^T & 0 \end{array} \right ]~,
\]
where $0_{2k}$ is the column vector of $2k$ zeroes. We immediately obtain that 
\[
\sqrt{2} \tilde{A}_k^{\beta} =   Q_{2k+1} V_{C_{\beta, k}} Q_{2k+1}^T~,
\]
where
\[
C_{\beta, k} = \left [ \begin{array}{cccc} \chi_{k \beta} & & & \\
					      \chi_{\frac{2k-1}{2} \beta} & \chi_{(k-1) \beta} & & \\
						& \chi_{\frac{2k-3}{2} \beta} & \chi_{(k-2) \beta} & \\
						& & \ddots & \ddots \\
						& &  \chi_{\frac{\beta}{2}}& 0 \end{array} \right ]
\]
with equality here being in the sense of distributions.

This is a second obstacle, as $C_{\beta, k}$ is not a $\beta$-Laguerre matrix, and has $0$ as a singular value. The latter part can easily be corrected by removing the last column of $C_{\beta, k}$ and creating a $(k+1) \times k$ matrix $\tilde{C}_{\beta, k}$.

We would now like to show that the $k \times k$ matrix $\tilde{L}_{\beta,k} = C_{\beta,k}^{T} C_{\beta,k}$ has the same eigenvalue distribution as the matrix $W_{\beta, k, a} = B_{\beta,k,a}^{T} B^{}_{\beta,k,a}$ with $a = \frac{2k+1}{4} \beta$.

\vspace{.2cm}

\noindent \emph{Notation.}
We will now make the following notational convention: in the below, any variable indexed by $i$ (e.g. $a_{i}$) will have distribution $\chi_{i \beta/2}$. Some indices will therefore be skipped.

\vspace{.2cm}

If we denote the entries of $C_{\beta, k}$ as follows:
\[
C_{\beta, k} = \left [ \begin{array}{cccc} b_{2k} & & & \\
				b_{2k-1} & b_{2(k-2)} & & \\
				& & \ddots & \ddots \\
				& & & b_{1} \end{array} \right ]~,
\]
then 
\[
\tilde{L}_{\beta,k} = \left [ \begin{array}{cccc} b_{2k}^2 + b_{2k-1}^2 & b_{2k-1} b_{2k-2} & & 
\\
b_{2k-1} b_{2k-2} & b_{2k-2}^2 + b_{2k-3}^2 & b_{2k-4}b_{2k-3} & \\
& & \ddots & \ddots \\
& & b_{2} b_{3} & b_1^2 + b_2^2 \\ \end{array} \right ]
\]
while at the same time, if we denote the entries of $B_{\beta,k,a}$ by
\[
B_{\beta,k,a} = \left [ \begin{array}{cccc} a_{2k+1} & & & \\
					    a_{2k-2} & a_{2k-1} & \\
					    & & \ddots & \ddots \\
					    & & a_{2} & a_3  \end{array} \right ]~,
\]
and
\[
W_{\beta,k, a} = \left [ \begin{array}{cccc} a_{2k+1}^2 + a_{2k-2}^2 & a_{2k-1} a_{2k-2} & & 
\\
a_{2k-1} a_{2k-2} & a_{2k-1}^2 + a_{2k-4}^2 & a_{2k-4}a_{2k-3} & \\
& & \ddots & \ddots \\
& & a_{2} a_{3} & a_3^2 \\ \end{array} \right ]
\]

Note that, with the notational convention adopted above, the marginals of the entries of $\tilde{L}_{\beta,k}$ and $W_{\beta,k, a}$ are the same, since independent chi-square variables add to a chi-square variable. 

\begin{claim}
The Choleski factorization of the matrix $\tilde{L}_{\beta,k}$ yields a matrix whose distribution is the same as $W_{\beta,k, a}$. 
\end{claim}

\begin{proof} Note that $\tilde{C}_{\beta,k}$ is not the Choleski factor of $\tilde{L}_{\beta,k}$, because it is $(k+1) \times k$ instead of $k \times k$.

We will prove that if we solve the system of equations
\begin{eqnarray*}
x_{2i+1}^2 + x_{2i-2}^2 &=& b_{2i}^2 + b_{2i-1}^2~, ~~\mbox{for}~i = 1,2,\ldots,k~, \\
x_{2i+1}x_{2i} & = & b_{2i+1} b_{2i}~, ~~\mbox{for}~i=1,2,\ldots,k-1~,
\end{eqnarray*}
with $(b_1^2, \ldots, b_{2k+1}^2)$ being independent chi-squared variables of parameter $(\beta/2, \ldots, (2k+1) \beta/2)$, then $(x_2^2, \ldots, x_{2k-1}^2) \sim (b_2^2, \ldots, b_{2k-1}^2)$. Moreover, we will obtain as a bonus that $x_{2k+1}^2$ is chi-square distributed, independently of all others, with parameter $\frac{2k+1}{2} \beta$. 

We first need the well-known lemma below.

\begin{lemma} \label{stiut} If $x \sim \chi^2_r$ and $y \sim \chi^2_s$, and $x,y$ are independent, then $z = \frac{x}{x+y}$ is distributed like $Beta(r,s)$, and $z$ is independent of $(x+y)$. Moreover, if $w \sim \chi_{r+s}^2$ independently of $x$ and $y$, then $wz \sim \chi^2_r$, $w(1-z) \sim \chi^2_s$, and $wz$ is independent of $w(1-z)$.
\end{lemma}

First, we find $x_2^2$ and $x_3^2$:
\begin{eqnarray*}
x_3^2 & = & b_1^2+b_2^2~, \\
x_2^2 & = & b_3^2 \frac{b_2^2}{b_1^2+b_2^2}~.
\end{eqnarray*}
It follows immediately from Lemma \ref{stiut} that $(b_3^2 - x_2^2,~x_2^2, x_3^2) \sim (b_1^2, b_2^2, b_3^2)$.

Given $(x_2^2, \ldots, x_{2i-1}^2)$ and $b_{2i-1}^2 - x_{2i-2}^2 $, we can obtain \begin{eqnarray*}
x_{2i+1}^2 & = & b_{2i}^2 + b_{2i-1}^2 - x_{2i-2}^2~ ~\mbox{and}~\\
x_{2i}^2 & =  & \frac{b_{2i}^2 b_{2i+1}^2}{x_{2i+1}^2}~~.
\end{eqnarray*}

Assume now that $$(b_{2i-1}^2 - x_{2i-2}^2, x_2^2, x_3^2, \ldots, x_{2i-1}^2) \sim (b_1^2, b_2^2, b_3^2, \ldots, b_{2i-1}^2)$$ for some $i \geq 2$. From the formulae above and Lemma \ref{stiut}, one sees that $(x_{2i}^2, x_{2i+1}^2) \sim (b_{2i}^2, b_{2i+1}^2)$ and that they are independent of all $x_{j}^2$ with $j \leq 2i-1$. Furthermore, $b_{2i+1}^2$ is independent of  all $x_{j}^2$ with $j \leq 2i+1$. 

Altogether, this yields $$(b_{2i+1}^2 - x_{2i}^2, x_2^2, x_3^2, \ldots, x_{2i+1}^2) \sim (b_1^2, b_2^2, b_3^2, \ldots, b_{2i+1}^2)~,$$ 
one can easily see that $x_{2k+1}^2 \sim b_{2k}^2+ b_1^2$ and that it is independent of all other $b$'s, and thus of all other $x$'s.

Thus the claim is proved by induction. \end{proof}

It remains to conclude that, since the Choleski factorization of the matrix $\tilde{L}_{\beta,k}$ yields a matrix whose distribution is the same as $W_{\beta,k, a}$, Theorems \ref{one} and \ref{two} are true for $n = 2k+1$.
\qed

\section{Sturm sequences and Pr\"ufer phases} \label{apps}
The characteristic polynomial $P_n(\lambda)$ for $iT$ with $T$ the anti-symmetric tridiagonal matrix
(\ref{def_trid}) is identical to the characteristic polynomial for the symmetric tridiagonal
matrix $T_s$ obtained from $T$ by removing the minus signs below the diagonal. This
can be seen from the three term recurrence (\ref{pbp}). Hence some fundamental results
applying to the characteristic polynomials of real symmetric tridiagonal matrices also
apply to $P_n(\lambda)$.

One such result relates to $N(\mu)$, the number of eigenvalues less than $\mu$. With
$r_i := - P_i(\mu)/P_{i-1}(\mu)$ $(i=1,\dots,n)$ the theory of Sturm sequences
(see e.g.~\cite{Wi65,ACE08}) tells us that $N(\mu)$ is equal to the number of negative values
in the sequence $\{r_i\}_{i=1,\dots,n}$ (assumming that $\mu$ is not an eigenvalue of $T$). In our anti-symmetric setting, there are precisely $[(n+1)/2]$ eigenvalues less than or equal to 0; since  $[(n+1)/2]$ is equal to the number of terms in $\{r_i\}_{i=1,3,\dots}$, giving the following result.

\begin{lemma}
Let $N^+(\mu)$ be the number of positive eigenvalues of $iT$ less than or
equal to $\mu$, and suppose $\mu$ is not an eigenvalue. We have that $N^+(\mu)$ is
equal to the number of negative values of $\{r_i\}_{i=2,4,\dots}$ minus the number
of positive values of $\{r_i\}_{i=1,3,\dots}$.
\end{lemma}

This result relates in turn to shooting eigenvectors $\mathbf{x} = (x_n,\dots,x_1)^T$ of
the symmetric tridiagonal matrix $T_s$. With $x_1$ and $\mu$ given, the shooting eigenvector
is specified as the solution of all but the first of the $n$ linear equations implied
by the matrix equation $(T_s - \mu \mathbb{I}) \mathbf{x} =  \mathbf{0}$. 
Further, with $x_{n+1}$ defined as the first component of $(T_s - \mu  \mathbb{I}) \mathbf{x}$,
we have that $x_i/x_{i-1} = - r_{i-1}/b_{i-1}$ as can be checked from the recurrence
(\ref{pbp}). 

Finally, we discuss the Pr\"ufer phases $\theta_i^\mu$ associated with the shooting
vectors of $T_s$ (see e.g.~\cite{JSS04}). For $2 \le i \le n$ these are specified in
terms of the characteristic polynomial by
\begin{equation}\label{1.5s}
\cot \theta_i^\mu = {1 \over b_{i-1}^2} {P_{i-1}(\mu) \over P_{i-2}(\mu)}
\end{equation}
where $b_n := 1$, together with the condition that
\begin{equation}\label{1.5as}
\theta_j^\mu \to - [j/2] \pi \qquad {\rm as} \quad \mu \to \infty
\end{equation}
(this is consistent with (\ref{1.5s}) as the RHS $\to \infty$ when
$\mu \to \infty$) and the requirement that $\theta_i^\mu$ be continuous in
$\mu$.

Differentiating (\ref{1.5s}) with respect to $\mu$ gives
$$
- \Big ( {d \theta_j^\mu \over d \mu} \Big ) {1 \over \sin^2  \theta_j^\mu} =
{1 \over b_{i-1}^2} \Big (
{P_{i-1}'(\mu) P_{i-2}(\mu) - P_{i-1}(\mu) P_{i-2}'(\mu) \over
(P_{i-2}(\mu))^2 } \Big ).
$$
But according to the Christoffel-Darboux summation formula the RHS is positive so we
recover the well know result (see e.g.~\cite{JSS04})  that $\theta_j^\mu$ 
$(j \ge 2)$ is a strictly decreasing function of $\mu$. Further, we see from
(\ref{1.5s}) and (\ref{1.5as}) that $\theta_i^\mu = \pi/2 + k \pi$, $k=1,2,\dots$ for
the $k$-th positive zero of $P_{i-1}(\mu)$, while $\theta_i^\mu = k \pi$,
$k=1,2,\dots$ for the $k$-th zero of $P_{i-2}(\mu)$. Since $\mu=0$ is a zero of
$P_j(\mu)$ for $j$ odd, it follows in particular that $\theta_{2j}^0= \pi/2$,
$\theta_{2j-1}^0 = 0$.  

In a significant advancement \cite{VV07} (see also \cite{KS06}), the Pr\"ufer phases
associated with the tridiagonal matrix (\ref{25}), (\ref{3.1}) relating to
the Gaussian $\beta$-ensemble, have been shown to
satisfy a stochastic differential equation. An analogous study of the  Pr\"ufer phases of the
present anti-symmetric Gaussian $\beta$-ensemble awaits investigation.

\section*{Acknowledgements}

We wish to thank one anonymous referee of a previous version of this manuscript for suggesting to us the proof for Method III, $n$ odd.

The work of PJF was supported by the Australian Research Council. We thank Raj Rao and
Alan Edelman for inviting us both to be part of the Workshop on Stochastic Eigen-Analysis
at FoCM'08, and so making this collaboration possible. ID would also like to thank Alan Edelman for introducing her to the ``perfect shuffle'' permutation.

\section*{Appendix}
\renewcommand{\theequation}{A.\arabic{equation}}
\setcounter{equation}{0}
As mentioned below (\ref{sapte}), the strategy of our proof of Lemma \ref{jacobian} was first used in
\cite{DE02} to compute the Jacobian for the change of variables from the elements of a general real
symmetric matrix to its eigenvalues and first component of its eigenvectors. Subsequently, a direct
approach to the computation of this Jacobian was given \cite{FR04}. In this appendix, we will show
how this direct approach can be adapted to provide an alternative 
proof for Lemma \ref{jacobian}.

Suppose first that $n$ is even. The starting point is the identity (\ref{pcp}) as it applies
to (\ref{def_trid}), with the LHS rewritten to read
according to Cramer's rule. Thus, after further minor manipulation, we have
\begin{equation}\label{A.pcp}
\Big ( ( {\mathbb I}_n - \lambda i T)^{-1} \Big )_{11} = \sum_{j=1}^{n/2} {2 q_j^2 \over 1 - \lambda^2 \lambda_j^2}.
\end{equation}
By equating successive powers of $\lambda$ on both sides we deduce
\begin{eqnarray}\label{a.1}
&&1 = \sum_{j=1}^{n/2} 2 q_j^2, \qquad b_{n-1}^2 = \sum_{j=1}^{n/2} 2 q_j^2 \lambda_j^2, \qquad 
* + b_{n-2}^2 b_{n-1}^2 = \sum_{j=1}^{n/2} 2 q_j^2 \lambda_j^4, \nonumber \\
&&* + b_{n-3}^2 b_{n-2}^2 b_{n-1}^2 = \sum_{j=1}^{n/2} 2 q_j^2 \lambda_j^6, \qquad
 \cdots, * + \prod_{i=1}^{n-1} b_i^2 = \sum_{j=1}^{n/2} 2 q_j^2 \lambda_j^{2n-2}
\end{eqnarray}
where the * denotes terms which have already appeared on the LHS of preceding equations. In
particular, the set of equations (\ref{a.1}) is triangular in $b_{n-1}, b_{n-2},\dots,b_1$, and the
first of these equations implies
\begin{equation}\label{A.qq}
q_{n/2} dq_{n/2} = - \sum_{j=1}^{n/2} q_j dq_j.
\end{equation}
Taking differentials of the remaining equations, substituting for $q_{n/2} dq_{n/2}$, and taking
the wedge product of both sides shows
\begin{equation}\label{A.a.2}
\prod_{j=1}^{n-1} b_j^{2j-1} d \mathbb{b} = {2^{n-1} \over q_{n/2} } \prod_{j=1}^{n/2} q_j^3
\det \Big [ [\lambda_k^{2j} - \lambda_{n/2}^{2j}]_{j=1,\dots,n-1 \atop k=1,\dots,n/2-1} \: \:
[j \lambda_k^{2j-1}]_{j=1,\dots,n-1 \atop k=1,\dots,n/2} \Big ].
\end{equation}
Here, to obtain the LHS essential use has been made of the triangular structure, and we
have written
\begin{equation}\label{A.a.2a}
d \mathbf{b} := \wedge_{j=1}^{n-1} d b_j, \qquad
d \mathbf{\lambda} := \prod_{j=1}^{n/2} d \lambda_j, \qquad
d \mathbf{q} := \prod_{j=1}^{n/2} dq_j.
\end{equation}
According to \cite[Proposition 2.1]{FR04}, up to a sign the determinant in (\ref{A.a.2})
evaluates to
$$
\prod_{j=1}^{m/2} \Big ( \Delta(\lambda^2) \Big )^4,
$$
so after making use too of Lemma \ref{vander} we have
$$
d \mathbf{b} = {1 \over 2 q_{n/2}} {\prod_{j=1}^{n-1} b_j \over \prod_{j=1}^{n/2} q_j \lambda_j}
d \mathbf{\lambda} \wedge d \mathbf{q}.
$$
Noting that $d \mathbf{q}/2 q_{n/2} = {\rm d} q$, as follows from the meaning of ${\rm d}q$ below
(\ref{distr}) and a simple scaling, we read off the Jacobian ${\mathcal J}$ as specified in
the first case of Lemma \ref{jacobian}. 

For $n$ odd, rewriting (\ref{pcp}) as in (\ref{A.pcp}) gives
$$
\Big ( ( {\mathbb I}_n - \lambda i T)^{-1} \Big )_{11} = \sum_{j=1}^{(n-1)/2} {2 q_j^2 \over 1 - \lambda^2 \lambda_j^2}
+ c^2.
$$
We see that the first equation in (\ref{a.1}) needs to be modified to read
$$
1 = \sum_{j=1}^{(n-1)/2} 2 q_j^2 + c^2,
$$
and the remaining equations remain the same except that the upper terminal in the summation
on the RHS must have $n/2$ replaced by $(n-1)/2$. In particular, the variable $c$ does not
appear in any other equation, and we obtain in place of (\ref{A.a.2})
\begin{equation}\label{A.a.24}
\prod_{j=1}^{n-1} b_j^{2j-1} d \mathbf{b} = 2^{n-1} \prod_{j=1}^{(n-1)/2} q_j^3
\det \Big [ [\lambda_k^{2j} ]_{j=1,\dots,n-1 \atop k=1,\dots,(n-1)/2} \: \:
[j \lambda_k^{2j-1}]_{j=1,\dots,n-1 \atop k=1,\dots,(n-1)/2} \Big ]
\end{equation}
where $d \mathbf{\lambda}$ and $d \mathbf{q}$ are as in (\ref{A.a.2}) but with the upper terminals
in the products replaced by $(n-1)/2$. The argument used to establish \cite[Proposition 2.1]{FR04}
shows that up to a sign the determinant is equal to
$$
\prod_{j=1}^{n/2} \lambda_j^5  \Big ( \Delta(\lambda^2) \Big )^4.
$$
Hence, after substituting in (\ref{A.a.24}) and using  Lemma \ref{vander} we obtain for the
Jacobian
$$
{\prod_{i=1}^{n-1} b_i \over c^2 \prod_{j=1}^{(n-1)/2} q_j \lambda_j}.
$$
This agrees with the second case of Lemma \ref{jacobian}, after noting that
$d \mathbf{q}/c = {\rm d} q$, with ${\rm d} q$ as specified below (\ref{distr}).


\providecommand{\bysame}{\leavevmode\hbox to3em{\hrulefill}\thinspace}
\providecommand{\MR}{\relax\ifhmode\unskip\space\fi MR }
\providecommand{\MRhref}[2]{%
  \href{http://www.ams.org/mathscinet-getitem?mr=#1}{#2}
}
\providecommand{\href}[2]{#2}

\end{document}